\documentclass[aps, prd, superscriptaddress, preprintnumbers, floatfix]{revtex4}

\usepackage{graphicx,amsfonts,amsmath,amssymb,amstext}
\usepackage{float,wrapfig}
\usepackage{subfigure, psfrag}
\usepackage{dsfont}
\usepackage{color}
\usepackage{multirow}
\usepackage{hyperref}
\hypersetup{
	colorlinks = true,
	linkcolor = blue,
	citecolor = blue
}
\usepackage{bm}

\newcommand{\be}{\begin{equation}}
\newcommand{\ee}{\end{equation}}
\newcommand{\ba}{\begin{eqnarray}}
\newcommand{\ea}{\end{eqnarray}}

\definecolor{purple}{rgb}{0.8,0,0.6}
\definecolor{darkgreen}{rgb}{0.00,0.6,0.00}

\renewcommand{\arraystretch}{2}

\extrafloats{100}

\begin{document}

\title{Bose-Einstein condensate of Dirac magnons: Pumping and collective modes}
\date{January 4, 2021}

\author{P.~O.~Sukhachov}
\email{pavlo.sukhachov@su.se}
\affiliation{Nordita, KTH Royal Institute of Technology and Stockholm University, Roslagstullsbacken 23, SE-106 91 Stockholm, Sweden}
\affiliation{Department of Physics, Yale University, New Haven, CT 06520, USA}

\author{S.~Banerjee}
\email{saikat.banerjee@physik.uni-augsburg.de}
\affiliation{Theoretical Physics III, Center for Electronic Correlations and Magnetism, Institute of Physics, University of Augsburg, 86135 Augsburg, Germany}

\author{A.~V.~Balatsky}
\email{balatsky@gmail.com}
\affiliation{Department of Physics, University of Connecticut, Storrs, CT 06269, USA}
\affiliation{Nordita, KTH Royal Institute of Technology and Stockholm University, Roslagstullsbacken 23, SE-106 91 Stockholm, Sweden}

\begin{abstract}
We explore the formation and collective modes of Bose-Einstein condensate of Dirac magnons (Dirac BEC). While we focus on two-dimensional Dirac magnons, an employed approach is general and could be used to describe Bose-Einstein condensates with linear quasiparticle spectrum in various systems. By using a phenomenological multicomponent model of pumped boson population together with bosons residing at Dirac nodes, the formation and time evolution of condensates of Dirac bosons is investigated.
The condensate coherence and its multicomponent nature are manifested in the Rabi oscillations whose period is determined by the gap in the spin-wave spectrum. A Dirac nature of the condensates could be also probed by the spectrum of collective modes. It is shown that the Haldane gap provides an efficient means to tune between the gapped and gapless collective modes as well as controls their stability.
\end{abstract}

\maketitle

\section{Introduction}
\label{sec:Introduction}

Discrete parity and time-reversal symmetries as well as non-Bravais lattices lead to relativistic-like Dirac energy spectrum in what is called fermionic Dirac and Weyl semimetals~\cite{Katsnelson:rev-2007,Geim-Novoselov:rev-2007,CastroNeto-Geim:rev-2009,Geim:rev-2009,Katsnelson:book-2012,Hasan-Kane:rev-2010,Qi-Zhang:rev-2010,Bernevig:book-2013,Franz:book-2013,Wehling-Balatsky:rev-2014,Yan-Felser:2017-Rev,Hasan-Huang:rev-2017,Armitage-Vishwanath:2017-Rev}. The same attributes can be arranged in the case where the lattice is populated with bosons. Thus, one can easily extend the notion of Dirac and Weyl materials to include bosonic Dirac materials that host multicomponent elementary bosonic excitations described effectively by the Dirac or Weyl Hamiltonian.
Dirac bosons were realized in photonic~\cite{Jacqmin-Amo:2014,Lu-Soljacic:2015,Wang-Jiang:2016,Yang-Zhang:2018,Guo-Zhang:2018,Yang-Chen:2019} and phononic~\cite{Xiao-Chan:2015,Serra-Garcia-Huber:2018,Cai-Liu:2020} crystals and magnets~\cite{Pershoguba-Balatsky:2018,Chen-Dai:2018,Lee-Hammel:2020}.
Honeycomb arrays of superconducting grains were also predicted to host bosonic Dirac excitations~\cite{Banerjee-Balatsky:2016}. Photonic and phononic nodal excitations are seen in specifically crafted metamaterials that reveal Dirac points when illuminated by electromagnetic or acoustic waves. For example, a linear graphene-like dispersion relation of polaritons was observed in the photoluminescence spectrum of a honeycomb micropillar lattice in Ref.~\cite{Jacqmin-Amo:2014}.

It is known that a Dirac-type quasiparticle spectrum is generated for localized magnetic moments on a two-dimensional (2D) honeycomb lattice~\cite{Fransson-Balatsky:2016,Pershoguba-Balatsky:2018}.
The corresponding magnon Dirac materials~\cite{Fransson-Balatsky:2016} such as van der Waals crystals Cr$_2$Ge$_2$Te$_6$ ~\cite{Gong-Zhang:2017} and the transition metal trihalides family $AX_3$ (where $A=\mbox{Cr}$ and $X=\mbox{F,Cl,Br,I}$), e.g., CrI$_3$~\cite{Huang-Xu:2017,Pershoguba-Balatsky:2018,Chen-Dai:2018,Lee-Hammel:2020} received attention recently. These materials are layered ferromagnets with typical ferromagnetic transition in the temperature range of 10-50~K. Transition metal trihalides were known for a long time~\cite{Gossard-Remeika:1961,Davis-Narath:1964} and can be viewed as a magnetic analog of $ABC$ stacked graphite.
In three dimensions (3D), a magnonic Weyl state was predicted in pyrochlores~\cite{Li-Chen:2016,Mook-Mertig:2016,Owerre:2018}. Linear crossing in the magnon spectrum was indeed observed in the antiferromagnet Cu$_3$TeO$_6$~\cite{Yao-Li:2018} and the multiferroic ferrimagnet Cu$_2$OSeO$_3$~\cite{Zhang-Mokrousov:2020}.

A wide range of properties was observed within this emergent class of Dirac magnon materials. We mention
interaction effects~\cite{Pershoguba-Balatsky:2018,Banerjee-Balatsky:2020} that renormalize the slope of the dispersion, topological properties~\cite{Zhang-Li:2013,Mook-Mertig:2014,Owerre:2016,Nakata-Loss:2017,Kovalev-Li:2017,Owerre:2017,Zhang-Mokrousov:2020,Wang-Troncoso:2020}, strain effects~\cite{Ferreiros-Vozmediano:2018,Nayga-Vojta:2019,Liu-Shi:2020}, and non-Hermitian dynamics~\cite{McClarty-Rau:2019}.
While some properties of fermionic and bosonic Dirac materials are similar, there are, of course, crucial differences. The key is the difference in particle statistics:  there is no Pauli principle for bosons. Therefore, no Fermi surface exists in bosonic Dirac materials.

For the excitations like magnons or polaritons, the equilibrium chemical potential is zero because these excitations are absent in the ground state in equilibrium.
Thus, no Bose-Einstein condensate (BEC) of magnons is possible in equilibrium. Nevertheless, one can ask if it possible to create the coherent states or even Bose-Einstein condensate of Dirac bosons and what would be the proper description? We propose that the answer to this question is positive and suggest a realization of coherent Dirac bosons and condensates by creating steady-state nonequilibrium population at the Dirac nodes. Such condensates could naturally be viewed as {\em Dirac BECs}: they would have multiple component and the nodal spectrum  described by an effective Hamiltonian resembling that of Dirac (quasi-)particles.

Dirac nodes are usually located at higher energies where, unlike fermions, a nonvanishing population cannot be achieved via equilibrium doping or gating.
Therefore, one would need to produce excitations (i.e., pump the material) to populate the Dirac points, e.g., see the discussion on pumping for magnons below.
(For population of the Dirac nodes via light pulses for fermions, see Ref.~\cite{Pertsova-Balatsky:2020}.)

Bosonic nature of excitations and pumping allow for an accumulation of Dirac bosons in the state with the same energy and ultimately opens up the possibility of Bose-Einstein condensation at the Dirac point. This uncovers a treasure trove of various nontrivial effects discussed for conventional BECs~\cite{Dalfovo-Stringari:1999-rev,Ozeri-Davidson:2005-rev,Pitaevskii-Stringari:book}. Recently, properties of BECs with Dirac points in the energy spectrum were studied in Refs.~\cite{Haddad-Carr:2011a,Haddad-Carr:2011b,Haddad-Carr:2015a,Haddad-Carr:2015b,Haddad-Carr:2015c,Haddad-Carr:2015d,Haddad-Carr:2015e} (see also a vast literature on the nonlinear Dirac equation, e.g., Refs.~\cite{Cuevas-Maraver-Lan:2016,Cuevas-Saxena:2018,Poddubny-Smirnova:2018}). Among the most interesting features, we notice various types of vortices and solitons~\cite{Haddad-Carr:2015a,Haddad-Carr:2015b,Haddad-Carr:2015c,Haddad-Carr:2015d}. As plausible systems for the realization of Dirac BECs, cold atoms and honeycomb optical lattices were proposed. In all of these studies, the prior equilibrium BEC was subjected to a perturbing potential to form the Dirac nodes.
To the best of our knowledge, however, the possibility and properties of Dirac magnon BECs were not investigated before.
Obviously, the ultimate proof of the existence of these condensates would be given by a successful experiment. Here, we lay out the theoretical framework that would facilitate the realizations of the Dirac BEC. The framework should be understood as a model suitable for Bose-Einstein condensates where bosons can be described by a relativistic-like Dirac equation.

Magnons represent a convenient and tested  platform for realizing quasiparticle coherent states and BEC in solids. For example, magnon BEC in yttrium–iron–garnet (YIG) films was recently experimentally observed and analyzed~\cite{Demokritov-Slavin:2006,Demidov-Slavin:2007,Dzyapko-Slavin:2007,Demidov-Slavin:2008,Nowik-Boltyk-Demokritov:2012,Serga-Hillebrands:2014,Bozhko-Hillebrands:2019,Mohseni-Pirro:2020,Mohseni-Pirro:2020b}.
(While we focus on magnons, the BEC of polaritons is also possible and was observed in, e.g., Refs.~\cite{Kasprzak-Dang:2006,Deng-Yamamoto:2010,Deveaud:2015,Sun-Nelson:2017,Proukakis-Littlewood:book-2017}.)
Among the most notable properties, we mention spatial coherence, supercurrents, and sound modes of the condensate. In order to achieve a magnon BEC, one needs to accumulate magnons in the minima of dispersion relation, which is usually achieved by a certain pumping protocol. Often used is the parametric pumping technique~\cite{Lvov:book}. The central process in this scheme is the  splitting of a photon of an external oscillating electromagnetic field into two magnons with arbitrary opposite momenta but with the half-frequency. These magnons subsequently accumulate at the minima of the dispersion and condense if the pumping intensity exceeds certain threshold. In contrast to the case of Dirac magnons, the spin-wave spectrum in YIG  is ``trivial", i.e., it does not exhibit any Dirac crossings and is well described by a conventional parabolic dispersion. Moreover, the vast literature investigating the properties of driven BEC in YIG relies on the standard single-component models for magnon BEC, which cannot be applied to Dirac quasiparticles.

For the case under consideration, we assume the nodal structure of spin excitations (Dirac nodes) and aim to describe the formation of long-lived and ultimately condensed magnons at Dirac nodes.
We derive  a minimal model for a Dirac BEC from the magnetic Hamiltonian, and  investigate the time evolution of pumped condensates and the properties of collective modes. While we focus on the case of 2D Dirac magnons, the model is general and can be applied for investigating various Dirac BECs. Since no Fermi surface exists for bosons, we apply pumping to generate a macroscopic occupancy of magnons and ultimately a magnon BEC. This process is phenomenologically described by a multicomponent system, which consists of nonlinear Gross-Pitaevskii equations for a Dirac magnon BEC and a rate equation for a pumped magnon population. As in the case of conventional magnon BECs, the condensate appears after reaching a certain pump power. The effective Haldane gap term, allowed by the Dirac nature of magnons, lifts the degeneracy between the magnon densities for different pseudospins (or sublattices). Furthermore, the gap leads to the Rabi oscillations of the magnon densities when the Josephson coupling terms are present. If experimentally observed, such oscillations would be a definitive signature of the condensate coherence. The nontrivial nature of Dirac BEC is manifested also in the spectrum of collective modes. In our analysis, we used the standard Bogolyubov approach and considered a homogeneous ground state. Interestingly, there are two possibilities for the latter: (i) magnon densities for both pseudospins are nonzero and (ii) one of the densities vanishes. The first state spontaneously breaks the rotational symmetry leading to two gapless Nambu-Goldstone modes. On the other hand, the second ground state is characterized by a gapped Higgs mode and a gapless mode. Depending on the model parameters, collective modes could be unstable. For example, the Haldane gap can be used to access different regimes for collective modes. To the best of our knowledge, the investigation of Dirac bosons, in particular, Dirac magnons, is in its infancy and a Dirac magnon BEC remains yet to be experimentally realized.
We hope that our findings will stimulate the corresponding experimental search and the development of the field of Dirac BECs.

The paper is organized as follows. We introduce the model and key notions in Sec.~\ref{sec:Model}. While we focus on the case of magnons, the model is general and could be used to describe various Dirac BECs. Pumping and time evolution of Dirac magnons are phenomenologically considered in Sec.~\ref{sec:phenomenology-pumping}. Section~\ref{sec:phenomenology-collective} is devoted to collective modes in the Dirac BECs. The results are discussed and summarized in Sec.~\ref{sec:Summary}. A schematic derivation of the free energy ansatz used in the main text is given in Appendix~\ref{sec:App-magnon-H}. A few non-trivial phases of the Dirac BEC are summarized in Appendix~\ref{sec:App-BEC-Class}. The additional results for the population dynamics at different interaction constants are presented in Appendix~\ref{sec:App-BEC-gab}. Appendix~\ref{sec:App-Josephson-omega} contains dispersion relations of collective modes for a nonzero Josephson coupling. The model is extended to include the magnon condensate at the $\Gamma$ point and the results for the population dynamics are presented in Appendix~\ref{sec:App-phenomenology-4component}. Through this study, we set $\hbar=k_{B}=1$.

\section{Model}
\label{sec:Model}

In this section, we present a phenomenological approach to Dirac BECs. The approach is similar to that for conventional BECs~\cite{Pitaevskii-Stringari:book} albeit it incorporates the pseudospin and a linear Dirac dispersion relation. We start with the following minimal ansatz for the free energy density (for a schematic derivation in the case of magnons, see Appendix~\ref{sec:App-magnon-H}):
\begin{eqnarray}
\label{phen-F-def}
F &=& (c_0-\mu) n_{\rm tot}+\Delta  \sum_{\zeta=\pm}(n_{a,\zeta}-n_{b,\zeta}) -iv \Psi^{\dag}(\tau_z\otimes\bm{\sigma}\cdot\bm{\nabla})\Psi \nonumber\\
&+& \sum_{\zeta=\pm} \left[\frac{1}{2}g_{a}n_{a,\zeta}n_{a,\zeta} + \frac{1}{2}g_{b}n_{b,\zeta}n_{b,\zeta} +g_{ab}n_{a,\zeta}n_{b,\zeta}\right] + \frac{1}{2}\sum_{\zeta=\pm} \left(u_{ab} \psi_{a,\zeta}^{*}\psi_{a,\zeta}^{*} \psi_{b,\zeta} \psi_{b,\zeta}+u_{ab}^{*} \psi_{b,\zeta}^{*}\psi_{b,\zeta}^{*} \psi_{a,\zeta} \psi_{a,\zeta}\right).
\end{eqnarray}
The four-component wave function of Dirac BEC is
\begin{equation}
\label{model-bispinor-def}
\Psi= \left(\psi_{a,+},\psi_{b,+},\psi_{b,-},\psi_{a,-}\right)^{T},
\end{equation}
where $\zeta=\pm$ denotes the $K$ or $K^{\prime}$ valley in 2D or chirality in 3D, $\psi_{a,\zeta}$ and $\psi_{b,\zeta}$ are the condensate wave functions for the different pseudospins, e.g., $A$ and $B$ sublattices for 2D hexagonal lattices, $\bm{\sigma}$ is the vector of the Pauli matrices in the pseudospin space, $\tau_z$ is the Pauli matrix in the valley space, $n_{i,\zeta}=\psi_{i,\zeta}^{*}\psi_{i,\zeta}$ with $i=a,b$ is the density for a certain pseudospin, $n_{\rm tot}=\sum_{\zeta=\pm}\left(n_{a,\zeta}+n_{b,\zeta}\right)$ is the total condensate density, $c_0$ is the position of the Dirac nodes in energy, and $\mu$ is the effective chemical potential of condensates.  Finally, the term quantified by $\Delta$ denotes the Haldane gap in 2D.  (Note that the term $\propto \Delta$ plays a different role in 3D where it shifts the Weyl nodes of opposite chiralities in momentum space and is called the chiral shift~\cite{Gorbar:shift-2009}. In what follows, however, we focus only on the 2D case.) Furthermore, we notice that the part of the free energy linear in the density is similar to that for other Dirac (quasi-)particles. In particular, the part with spatial derivatives resembles that for Dirac fermions. Moreover, there are two pseudospin or sublattice degrees of freedom. The fact that kinetic energy terms and Hilbert space of the excitations are very similar to the case of known Dirac excitations allows us to dub the condensate as the Dirac BEC.

In addition to the conventional for Dirac systems linear terms, we included also the nonlinear interaction terms quantified by constants $g_a$, $g_b$, and $g_{ab}$. While the former two correspond to the interaction strength of boson densities of the same pseudospin, the latter describes the inter-pseudospin interaction. In addition, the Josephson coupling term quantified by $u_{ab}$ is considered. This term is reminiscent to the spin-orbital coupling term in conventional spin-1 BECs~\cite{Lin-Spielman:2011}.
Cross-condensate Josephson terms describe the internal ``Josephson effect" where a pair of magnons of type $a$ converts into a pair of magnons of type $b$. As we will show below, the Josephson coupling term mixes the phases of the condensate wave functions corresponding to different pseudospins and might significantly affect the dynamics of the condensates.

Finally, we notice that several other terms are possible in the free energy, e.g., terms corresponding to the inter-valley mixing $\propto n_{a,-\zeta}n_{a,\zeta}$. For the sake of simplicity, we consider the effect of only a few parameters on Dirac condensates. A more detailed discussion is left for the future investigation.

\begin{figure*}[!ht]
\subfigure[]{\includegraphics[height=0.35\textwidth]{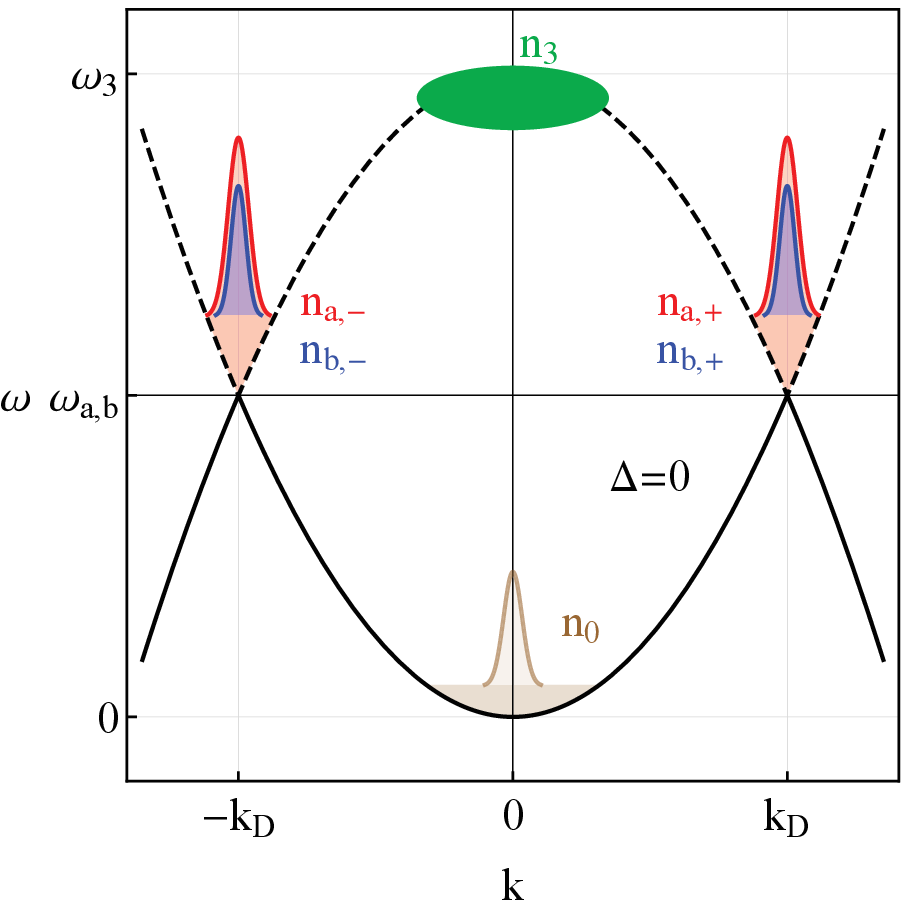}}
\hspace{0.1\textwidth}
\subfigure[]{\includegraphics[height=0.35\textwidth]{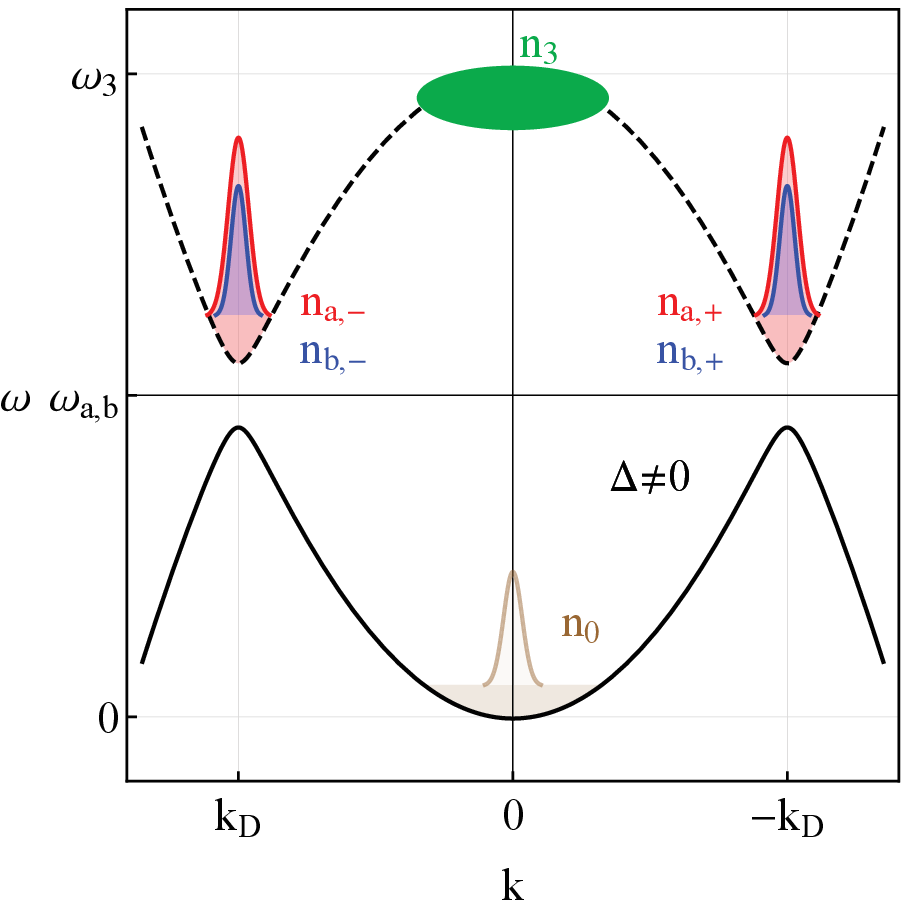}}
\caption{Schematic representation of the Dirac spectrum and BECs. The condensates of Dirac magnons occur at the Dirac points located at $k=\pm k_{\rm D}$ in momentum space. Panels (a) and (b) correspond to gapless $\Delta=0$ and gapped $\Delta\neq0$ spectra, respectively. Each of the condensates contains two components related to the pseudospin (sublattice) degree of freedom. The corresponding densities are denoted as $n_{a,\zeta}$ and $n_{b,\zeta}$ with $\zeta=\pm$. The density $n_3$ corresponds to the pumped population of bosons with energy $\omega_3$ (see Sec.~\ref{sec:phenomenology-pumping} for the description of the pumping mechanism). Finally, $n_0$ denotes the population at the $\Gamma$ point (see also Appendix~\ref{sec:App-phenomenology-4component}).
}
\label{fig:phen-spectrum-schematic}
\end{figure*}

The 2D Dirac dispersion relation (see Appendix~\ref{sec:App-magnon-H} for the discussion of the Dirac spin-wave dispersion relation) and BECs corresponding to different valleys and pseudospins are shown schematically in Fig.~\ref{fig:phen-spectrum-schematic}. Here, red and brown shaded regions correspond to the accumulated magnons (or other Dirac bosonic quasiparticles) and spikes represent condensates. Note that unlike fermionic systems, energy is calculated with respect to the bottom of the bands. Two Dirac points are situated at $k=\pm k_{\rm D}$ in momentum space and at $\omega_a=\omega_b=c_0$ in energy. As one can see from Fig.~\ref{fig:phen-spectrum-schematic}(b), the presence of the $\Delta$ term opens the gap in the spectrum.

By varying the free energy (\ref{phen-F-def}) with respect to the fields $\psi_{a,\zeta}^{*}$ and $\psi_{b,\zeta}^{*}$, we derive the following Gross-Pitaevskii equations for Dirac BECs at $A$ and $B$ sublattices:
\begin{eqnarray}
\label{phen-GPE-a-def}
i\partial_t \psi_{a,\zeta} &=& \left(c_0-\mu  +\Delta\right) \psi_{a,\zeta} -iv \left[\left(\zeta \partial_x -i\partial_y\right)\psi_{b,\zeta}\right]
+ g_{a} n_{a,\zeta} \psi_{a,\zeta} +g_{ab} n_{b,\zeta} \psi_{a,\zeta} +u_{ab} \psi_{a,\zeta}^{*}\left(\psi_{b,\zeta}\right)^{2},\\
\label{phen-GPE-b-def}
i\partial_t \psi_{b,\zeta} &=& \left(c_0-\mu -\Delta\right) \psi_{b,\zeta}  -iv \left[\left(\zeta \partial_x +i\partial_y\right)\psi_{a,\zeta}\right]
+ g_{b} n_{b,\zeta} \psi_{b,\zeta} +g_{ab} n_{a,\zeta} \psi_{b,\zeta} + u_{ab}^{*} \psi_{b,\zeta}^{*}\left(\psi_{a,\zeta}\right)^{2}.
\end{eqnarray}
These are nonlinear Dirac equations where bosonic fields have four degrees of freedom: two pseudospins and two valley or chirality indices.

It is interesting to notice the nontrivial topology of the multicomponent BEC. If all coupling terms are ignored, the corresponding state would correspond to the $U(1)_{a+} \times U(1)_{a-} \times U(1)_{b+} \times U(1)_{b-}$ symmetry. Therefore, one would expect the rich phase diagram with multiple phases that could emerge in Dirac BECs, where relative phases of these symmetries can be broken. For simplicity and to make an explicit case, we will assume that relative oscillations of the condensates in both valleys are suppressed and there are only two flavor symmetries, which are explicitly broken to $U(1)_a \times U(1)_b$. In other words, we consider a symmetric response of the valleys. Therefore, we fix $\zeta=+$ and omit the valley index when it is not necessary. Nevertheless, we mention some non-trivial phases of the non-interacting BECs in Appendix~\ref{sec:App-BEC-Class}.

Unlike fermions, where there is a Fermi surface that can be tuned to intersect Dirac points, bosons occupy the lowest energy state. Therefore, in order to investigate the properties of the Dirac BEC, one first needs to achieve a sufficient quasiparticle density at the Dirac points. As we discuss in the next section, Dirac Bose-Einstein condensation can be achieved in a steady state created by pumping.

\section{Generation of Dirac Bose-Einstein condensates via pumping}
\label{sec:phenomenology-pumping}

In this section, we consider pumping and time evolution of Dirac BECs. Since the pumping scheme and a few other details depends on the nature of Bose-Einstein condensates and, in general, are different for polaritons, magnons, Cooper pairs, etc., we focus on the case of 2D Dirac magnons. (For pumping and dynamical instabilities in fermionic Dirac matter, see, e.g., Ref.~\cite{Pertsova-Balatsky:2020}.) As we discussed in the introduction, Dirac magnons can be realized in, e.g., CrI$_3$ (see Refs.~\cite{Pershoguba-Balatsky:2018,Chen-Dai:2018,Lee-Hammel:2020}). Note that the Haldane gap for Dirac magnons can be generated via the pseudo-Zeeman effect (for a discussion of the strain-induced pseudo-Zeeman effect for fermions in graphene, see Ref.~\cite{Manes-Vozmediano:2013}). In addition, the relatively large gap in the Dirac spectrum could appear due to the Kitaev interaction. For a recent study in CrI$_3$, see, e.g., Ref.~\cite{Lee-Hammel:2020}.

\subsection{Three-component model}
\label{sec:phenomenology-3component}

To describe the pumped Dirac BEC, we introduce a phenomenological three-component model (or five-component if the valley index is taken into account). In addition to the magnon populations at the $A$ and $B$ sublattices for the $K$ point, we include the pumped magnon population with the energy $\omega_3$ (see also Fig.~\ref{fig:phen-spectrum-schematic}). The extension of this model to the case where the magnon BEC is formed at the $\Gamma$ point is presented in Appendix~\ref{sec:App-phenomenology-4component}. These populations could be achieved by the parametric pumping similarly to the case of conventional magnetic materials~\cite{Lvov:book} or any other suitable pumping scheme. The results below do not depend on the details of the pumping. The system reads as
\begin{eqnarray}
\label{phen-GPE-a-1}
&&i\partial_t \psi_{a} +i\Gamma_{a}\psi_{a}= \left(c_0-\mu  +\Delta\right) \psi_{a} -iv \left[\left(\partial_x -i\partial_y\right)\psi_{b}\right]
+ g_{a} n_a \psi_{a} +g_{ab} n_{b} \psi_{a} +u_{ab} \psi_{a}^{*}\left(\psi_{b}\right)^{2} +iP_{a3} \Gamma_{a} n_{3}\psi_{a},\\
\label{phen-GPE-b-1}
&&i\partial_t \psi_{b} +i\Gamma_{b}\psi_{b}= \left(c_0-\mu -\Delta\right) \psi_{b} -iv \left[\left(\partial_x +i\partial_y\right)\psi_{a}\right]
+ g_{b} n_b \psi_{b} +g_{ab} n_{a} \psi_{b} + u_{ab}^{*} \psi_{b}^{*}\left(\psi_{a}\right)^{2} +iP_{b3} \Gamma_{b} n_{3}\psi_{b},\\
\label{phen-GPE-pump-1}
&&\partial_t n_3 +\Gamma_3 n_3 = \Gamma_3 \tilde{P}(t) -\sum_{\zeta=\pm}\left(P_{3a} \Gamma_{a} n_{a,\zeta} +P_{3b} \Gamma_{b} n_{b,\zeta}\right) n_3.
\end{eqnarray}
Compared to Eqs.~(\ref{phen-GPE-a-def}) and (\ref{phen-GPE-b-def}), we have added the dissipation terms $i\Gamma_{a}\psi_{a}$ and $i\Gamma_{b}\psi_{b}$ that correspond to the magnon-phonon interactions. The terms with $P_{a3}$ and $P_{b3}$ describe to the inflow of magnons from the pumped population and act as the source terms for Dirac magnons. On the other hand, the terms with $P_{3a}$ and $P_{3b}$ correspond to the depletion of the pumped magnon population. The latter is described by the rate equation (\ref{phen-GPE-pump-1}) for the magnon density $n_3$. As in the case of the magnon BEC, $\Gamma_3$ denotes the decay rate. Finally, the term $\Gamma_3 \tilde{P}(t)$ describes the direct inflow of magnons caused by the pump. (Note that explicit dynamics of noncondensed magnons~\cite{Hahn-Kopietz:2020} at the Dirac nodes is not included in the above system. In the case under consideration, it is implicitly taken into account in the inflow terms.)
Therefore, the formation of the condensates is a multistep process where the pumping first creates a magnon gas which decays into the Dirac magnons.

For the sake of simplicity, we consider a simple step-like profile for the pump power
\begin{equation}
\label{phen-t-tP-def}
\tilde{P}(t) = \sqrt{P_{\rm max}}\, \theta(t-t_{\rm Begin}) \theta(t_{\rm End}-t).
\end{equation}
Here, the pump starts at $t=t_{\rm Begin}$ and ends at $t=t_{\rm End}$. Further, $P_{\rm max}$ is the power of the pumping field and $\theta(x)$ is the step function. The square-root dependence on the pump power is in agreement with the experimental data for YIG~\cite{Demidov-Slavin:2008,Demokritov-Slavin:2008} (see also Refs.~\cite{Rezende:2009,Rezende:2010} for the theoretical discussion). While only a simple step-like profile in Eq.~(\ref{phen-t-tP-def}) is considered in this work, the time profiles of magnon BECs can be modulated by applying more complicated time-dependent pump.

\subsection{Time evolution of Dirac magnons}
\label{sec:phenomenology-time}

In this subsection, we present the numerical results for the pumped magnon populations and discuss the role of the Haldane gap and the Josephson coupling terms. Unless otherwise stated, we use the following set of the model parameters:
\begin{eqnarray}
\label{phen-three-t-num-vars-be}
&&\Gamma_{a} = \Gamma_{b} =\Gamma_3 = 0.25\,t_0^{-1}, \quad c_0 = t_0^{-1},\quad \mu=0, \quad \Delta=0, \quad g_{a} =g_{b}= t_0^{-1},\quad g_{ab}=u_{ab} =0,\\
&&P_{3a}=P_{3b}=P_{a3}=P_{b3}=1, \quad t_{\rm Begin} = 10\,t_0, \quad t_{\rm End} = 200\,t_0.
\label{phen-three-t-num-vars-ee}
\end{eqnarray}
Here, $t_0$ is the characteristic interaction timescale. Since no experimental data are present for the Dirac magnon BEC, we use model parameters. In addition, we note that, in general, $P_{i3}\neq P_{3i}$ with $i=a,b$. For example, a strong disparity between the scattering efficiency of gaseous population to magnons at the bottom of the energy dispersion and vice versa is observed for YIG in Ref.~\cite{Bozhko-Hillebrands:2019}. Further, because of the space separation between the lattice sites, the inter-sublattice interaction constant $g_{ab}$ is expected to be much weaker than the intra-sublattice ones $g_a$ and $g_b$. Therefore, we focus on the case $|g_{ab}|\ll g_{a}, g_{b}$. The results for nonvanishing $g_{ab}$ are presented in Appendix~\ref{sec:App-BEC-gab}. Qualitatively, however, they are similar to the case $|g_{ab}|\ll g_{a}, g_{b}$. Finally, we note that the interaction constants $g_a$, $g_b$, and $g_{ab}$ are irrelevant for the condensate dynamics in the absence of the phase-mixing terms [e.g., the terms with $u_{ab}$ in Eqs.~(\ref{phen-GPE-a-1}) and (\ref{phen-GPE-b-1})]. This follows from the fact that the density dynamics is determined by the imaginary part of the Gross-Pitaevskii equations (\ref{phen-GPE-a-1}) and (\ref{phen-GPE-b-1}).

\begin{figure*}[!ht]
\centering
\subfigure[]{\includegraphics[height=0.28\textwidth]{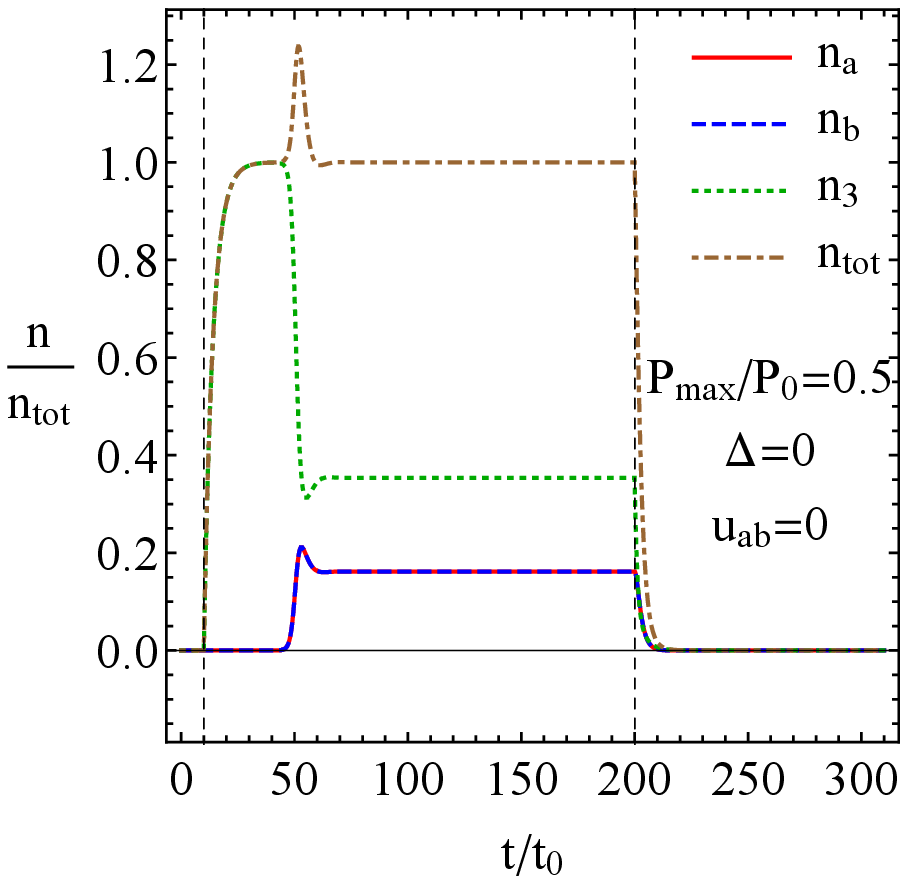}}
\hspace{0.02\textwidth}
\subfigure[]{\includegraphics[height=0.28\textwidth]{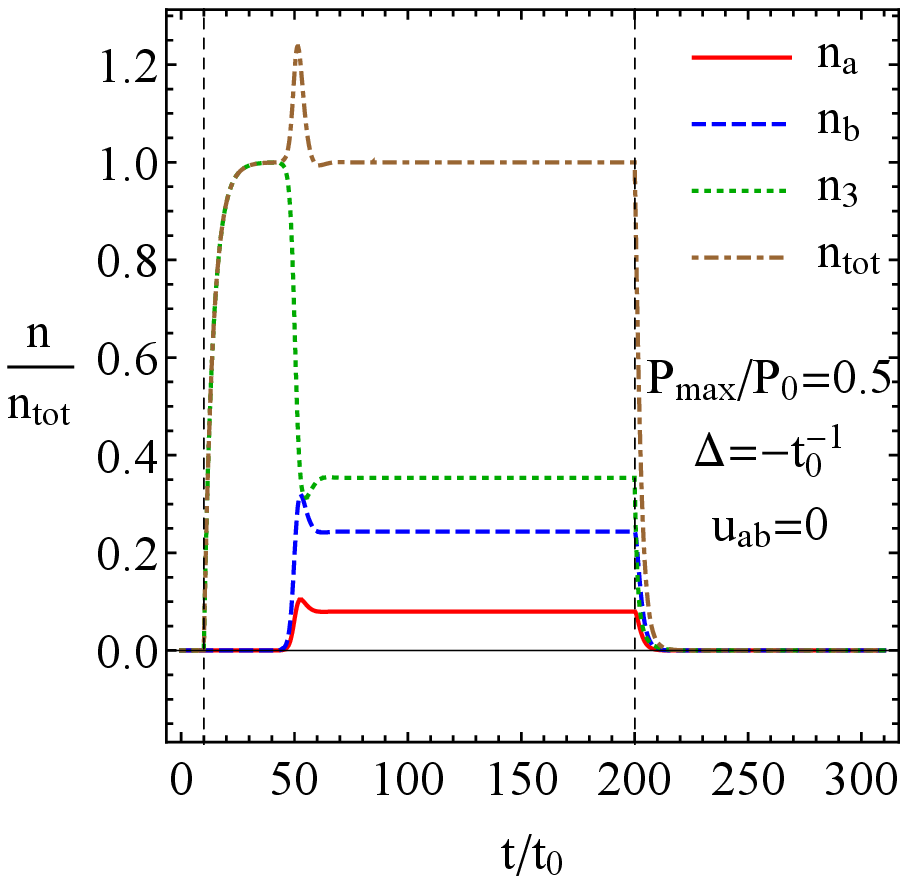}}
\hspace{0.02\textwidth}
\subfigure[]{\includegraphics[height=0.28\textwidth]{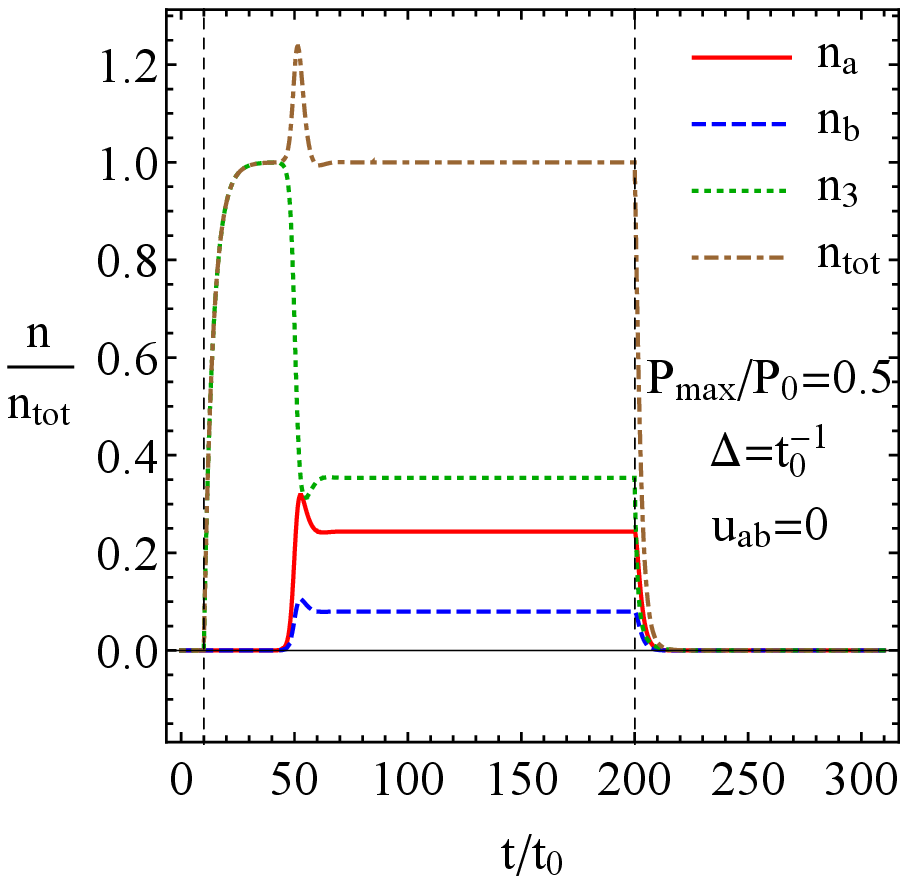}}
\caption{The dependence of the magnon densities on time $t$ at a few values of the Haldane gap $\Delta$ for $P_{\rm max}=0.5\,P_0$. The densities are normalized to the total magnon density at $t=t_{\rm End}$. Here, the values of parameters given in Eqs.~(\ref{phen-three-t-num-vars-be}) and (\ref{phen-three-t-num-vars-ee}) are used. In addition, $P_0$ is the characteristic strength of the pump ($P_0=t_0^{-2}$). The Haldane gap leads to the splitting of the magnon densities corresponding to different pseudospins.}
\label{fig:phen-three-t-P-Delta-I}
\end{figure*}

\begin{figure*}[!ht]
\centering
\subfigure[]{\includegraphics[height=0.35\textwidth]{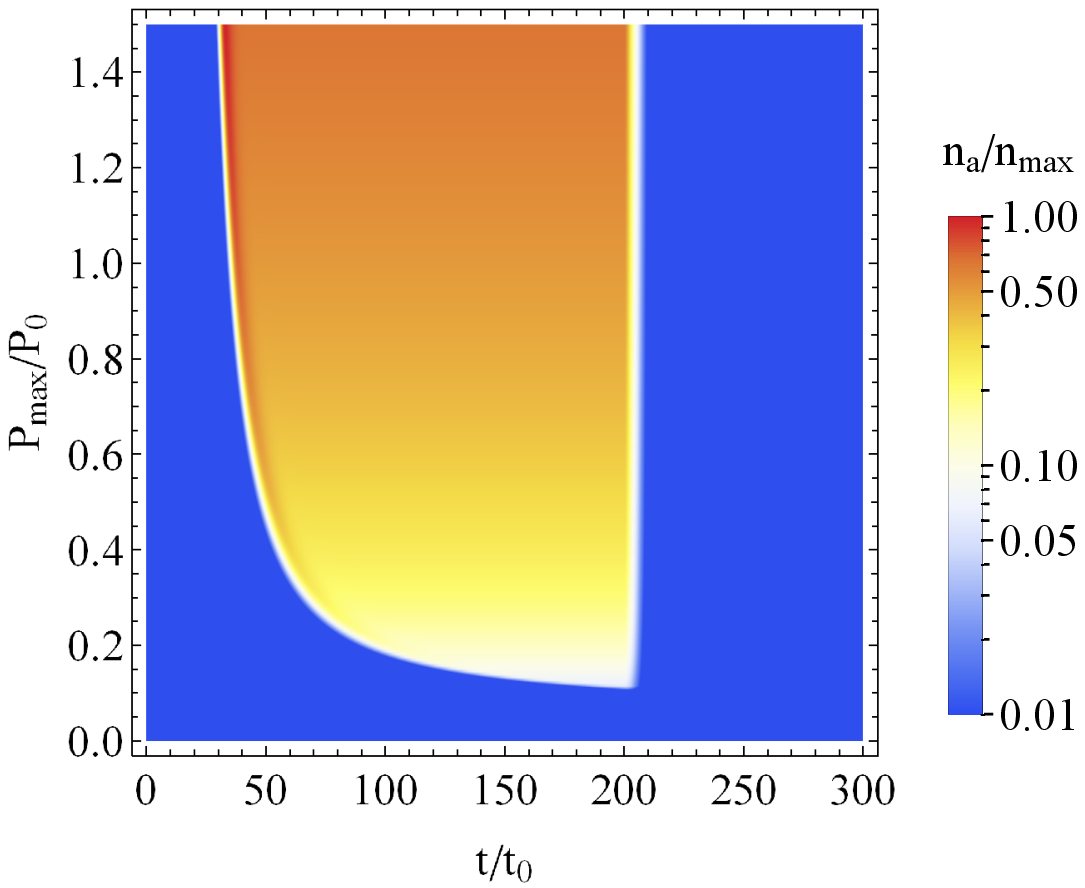}}
\hspace{0.1\textwidth}
\subfigure[]{\includegraphics[height=0.35\textwidth]{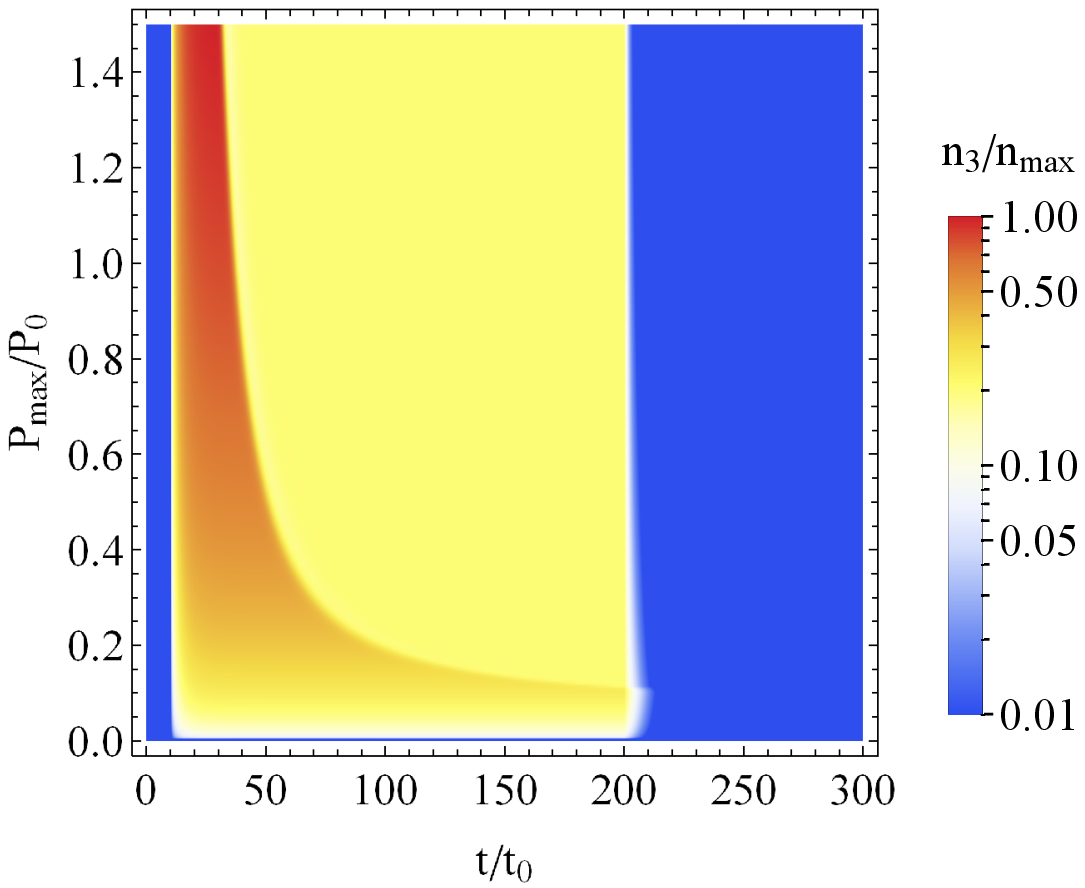}}
\caption{The dependence of the magnon densities (Dirac BEC in panel (a) and pumped population in panel (b)) on time $t$ and pump power $P_{\rm max}$. Here, the values of parameters given in Eqs.~(\ref{phen-three-t-num-vars-be}) and (\ref{phen-three-t-num-vars-ee}) are used. In addition, $P_0$ is the characteristic strength of the pump ($P_0=t_0^{-2}$). The pumped population demonstrates a box-shaped profile. On the other hand, the Dirac BEC forms much faster at larger powers.}
\label{fig:phen-three-t-P-I-density}
\end{figure*}

The time profiles of magnon densities are shown in Figs.~\ref{fig:phen-three-t-P-Delta-I} and \ref{fig:phen-three-t-P-I-density} for several values of the pump power. As one can see, the pumped magnon population $n_3$ quickly reaches its maximum value determined by the interplay of the pumping power and the decay rate. Magnon populations at the Dirac points occur later in time as well as require pump power to exceed a certain threshold (see Fig.~\ref{fig:phen-three-t-P-I-density}). It is interesting to note that the densities of the magnon BECs reach their maximum values at the onset of the profile. In addition, since we chose equal inflow and outflow terms in Eqs.~(\ref{phen-GPE-a-1}) through (\ref{phen-GPE-pump-1}), there is a noticeable depletion of the pumped population. After turning off the pump, both condensates and pumped population decay during the time $\tau\sim1/\Gamma$. It is notable, also, that the total number of magnons is not conserved and shows a spike when the BECs start to form. This could be qualitatively explained by the fact that the Dirac points accumulate more magnons during the rise phase than they can support in the steady state. These magnons are released during the equilibration phase.

Further, we investigate the effect of the Haldane gap $\Delta$. The corresponding results are shown in Fig.~\ref{fig:phen-three-t-P-Delta-I} at a few values of the gap for $P_{\rm max}=0.5\,P_0$. As one can see, the gap leads to different densities of magnon BECs at the $A$ and $B$ sublattices. Therefore, it can be used to effectively control the pseudospin distribution of Dirac magnon BECs.

\begin{figure*}[!ht]
\centering
\subfigure[]{\includegraphics[height=0.27\textwidth]{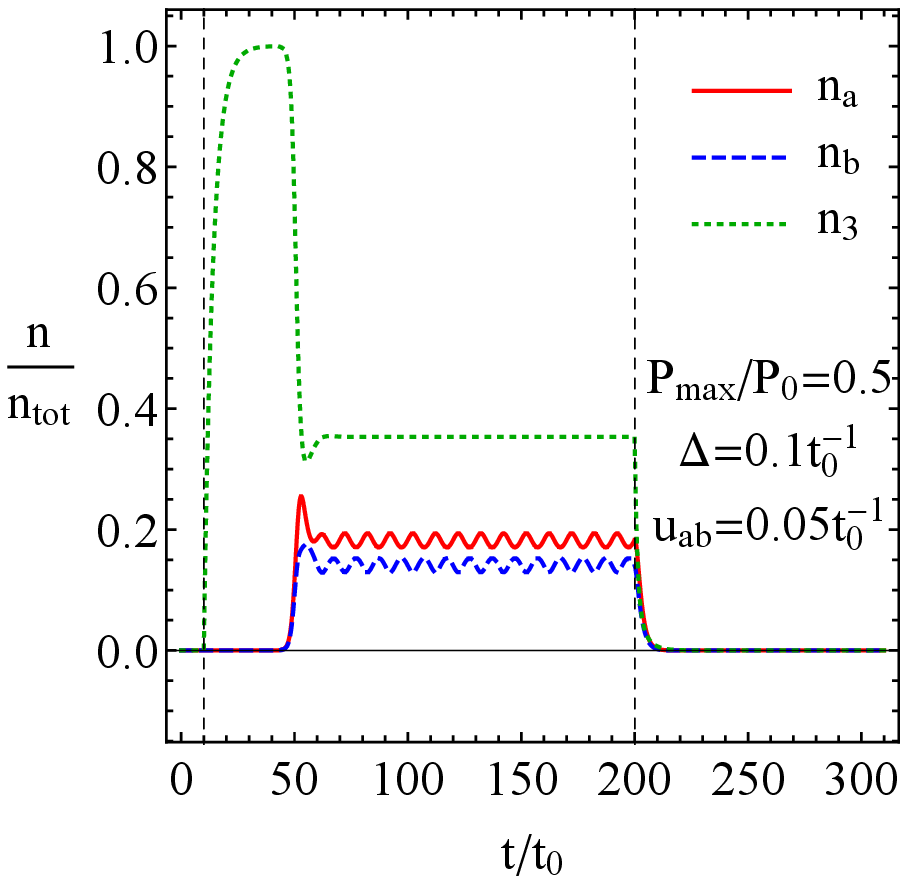}}
\hspace{0.02\textwidth}
\subfigure[]{\includegraphics[height=0.27\textwidth]{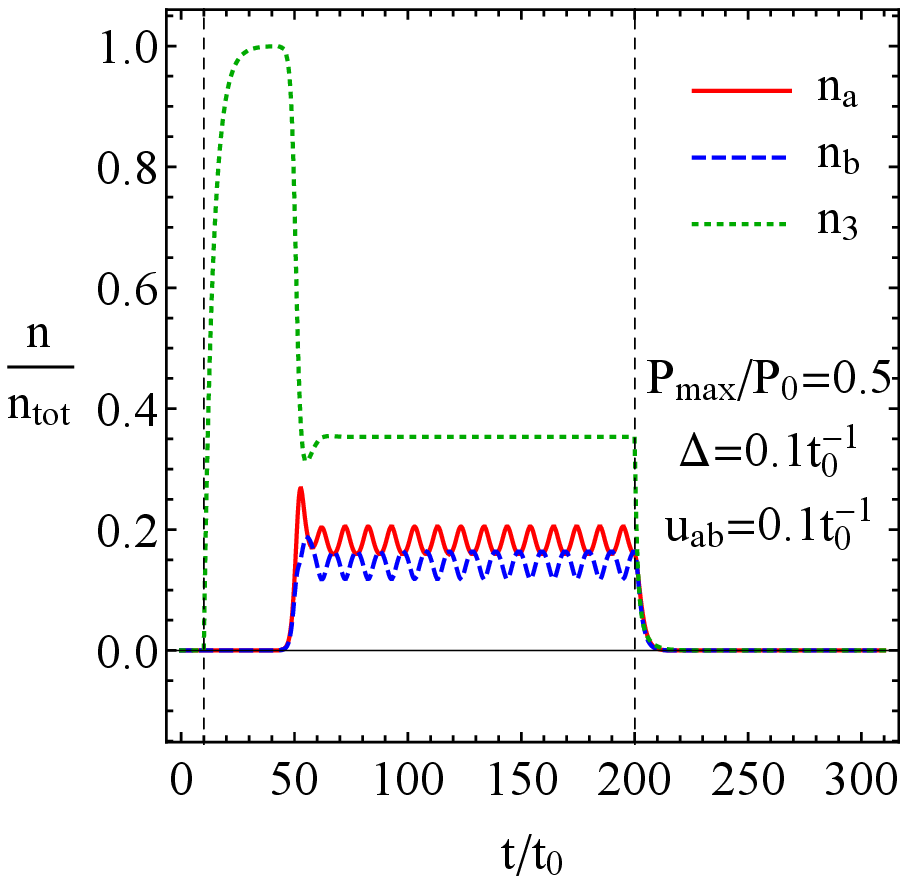}}
\hspace{0.02\textwidth}
\subfigure[]{\includegraphics[height=0.27\textwidth]{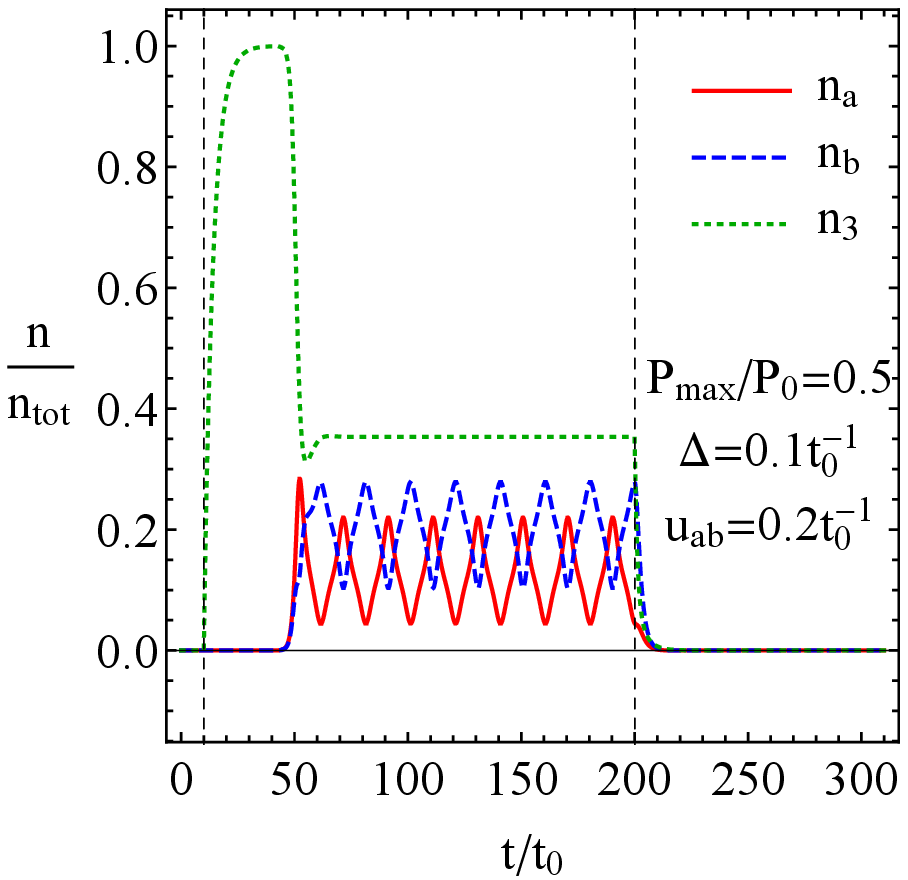}}
\caption{The dependence of the magnon densities on time $t$ at a few values of $u_{ab}$ for $P_{\rm max}=0.5\,P_0$ and $\Delta=0.1\,t_0^{-1}$. The densities are normalized to the total magnon density at $t=t_{\rm End}$. Here, the values of parameters given in Eqs.~(\ref{phen-three-t-num-vars-be}) and (\ref{phen-three-t-num-vars-ee}) are used. In addition, $P_0$ is the characteristic strength of the pump ($P_0=t_0^{-2}$). The Rabi oscillations, whose period is determined by $1/\Delta$ at small $u_{ab}$ and large $\Delta$ ($\Delta>0.1\,t_0^{-1}$ at $u_{ab}=0.2\,t_0^{-1}$), are clearly visible for the magnon BECs.}
\label{fig:phen-three-t-P-Josephson-uab1-uab}
\end{figure*}

\begin{figure*}[!ht]
\centering
\includegraphics[height=0.35\textwidth]{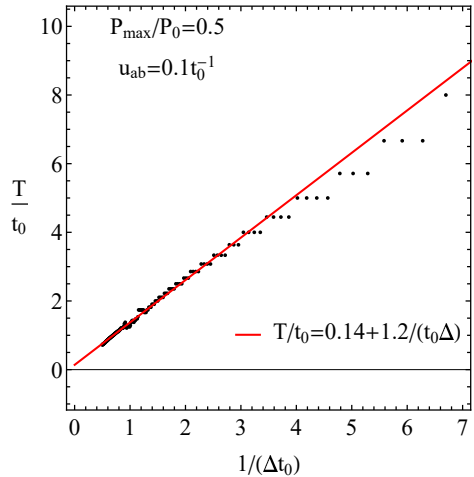}
\caption{The period of inter-condensate (Rabi) oscillations is shown by the black dots. The red line represents a linear fit. For large values of $\Delta$, we indeed observe the expected for the Rabi oscillations scaling for the period as inverse gap. On the other hand, for smaller values of the Haldane gap, the dynamic is more complicated and does not follow a simple linear behavior. We see a significant scatter in the period due to nonlinear dynamic at $\Delta\lesssim t_0^{-1}$. In this case, other interaction parameters also become important.}
\label{fig:phen-three-t-P-Josephson-period}
\end{figure*}

Finally, let us study the effect of the Josephson coupling term $u_{ab}$ on the time evolution of magnon densities. The corresponding results for a few values of $u_{ab}$ are shown in Fig.~\ref{fig:phen-three-t-P-Josephson-uab1-uab}. As one can see, the Josephson coupling term $\propto u_{ab}$ leads to the oscillations of the Dirac BEC. Oscillations of $n_a$ and $n_b$ always occur in the antiphase and the pumped magnon population remains constant. The observed oscillations are, in fact, the Rabi oscillations between magnons BECs with different pseudospins. To elucidate the nature of the inter-condensate oscillation and to show that they have the same scaling as the Rabi oscillation, we plot the corresponding period as a function of $\Delta$ in Fig.~\ref{fig:phen-three-t-P-Josephson-period}. The period indeed scales as $1/\Delta$ for large values of $\Delta$. The case of small $\Delta$ is more complicated, however. For example, the period becomes a nonmonotonic function and depends, e.g., on the value of $u_{ab}$ and other interaction parameters. A nontrivial dynamics becomes manifested at $\Delta\lesssim t_0^{-1}$.

Thus, the gap not only induces different magnon densities at the $A$ and $B$ sublattices, it can activate the Rabi oscillations between the condensates. We believe that the experimental observation of the Rabi oscillation of the magnon densities could serve as a definitive proof of their coherence.
In the systems with small distance between the Dirac nodes in the momentum space, such observations could be, in principle, done via the Brillouin light scattering technique used for conventional magnons in YIG. In the case of relatively large wave vectors, different techniques might be required to resolve magnons belonging to different Dirac nodes or sublattices. Among them, we mention time-resolved magneto-optic Kerr effect measurements and nitrogen-vacancy (NV) centers (see Refs.~\cite{Prananto-An:2020,Kruglyak-Grundler:2010} for reviews).

\begin{figure*}[!ht]
\centering
\includegraphics[height=0.35\textwidth]{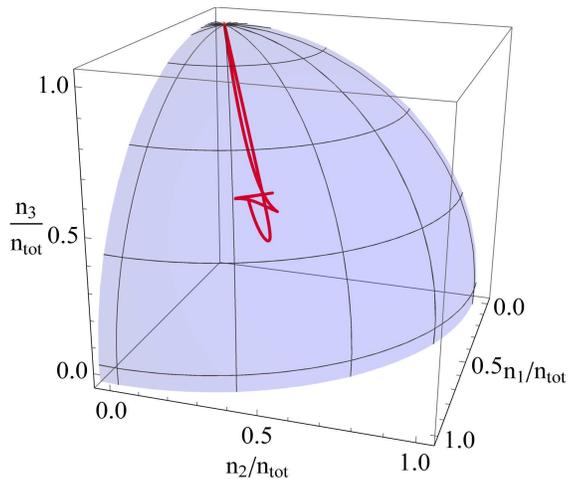}
\caption{The trajectory of the system in the density space $\left(n_{a},n_{b},n_3\right)/n_{\rm tot}$ is shown by the red line. Since the population densities are normalized to the total density, the dynamics is always constrained to the one-eight of a unit sphere. When the pump is turned on, only the pumped population $n_3$ contributes to the total density. Later, the condensates develop and the density vector start to oscillate closer to the equator. Finally, the condensates decay and the system returns to its initial state. Here, the values of parameters given in Eqs.~(\ref{phen-three-t-num-vars-be}) and (\ref{phen-three-t-num-vars-ee}) are used. In addition, we set $u_{ab}=0.1\,t_0^{-1}$ and $\Delta=0.025\,t_0^{-1}$. For the full time dependence, see the Supplemental Material~\cite{SM}.}
\label{fig:phen-three-Bloch}
\end{figure*}

The dynamics of the Dirac magnon populations can be presented by introducing the vector in the density space $\left(n_{a},n_{b},n_3\right)/n_{\rm tot}$. The tip of such vector is always confined to the one-eight of a unit sphere. The evolution of densities then resembles the dynamics of two-level quantum system whose pure states correspond to the surface of the Bloch sphere. The animation of dynamics of the Dirac magnon populations is given in the Supplemental Material~\cite{SM}. The snapshot after the pump is turned off ($t=300\,t_0$) is shown in Fig.~\ref{fig:phen-three-Bloch}.

\section{Collective modes}
\label{sec:phenomenology-collective}

In the previous section, we considered pumping of uniformly distributed magnons. To provide a deeper insight into the Dirac nature of magnon condensate and uncover properties related to the linear spectrum, one should consider coordinate-dependent probes. Collective modes are, perhaps, among the most well-known, powerful, and versatile of them.

Before we proceed with more detailed analysis, let us briefly discuss the structure of collective modes. There are two condensate densities $n_a$ and $n_b$. Hence we would expect two sound modes where the coupling is independent of the relative phase. Two linear modes will remain linear with renormalized velocities once the interaction is turned on. This is what we indeed find in the regime $n_a \neq 0$ and  $n_b \neq 0$. However, we also find the state where the condensate breaks the pseudospin symmetry and one of the components vanishes in the ground (or, more precisely, quasi-ground) state, e.g., $n_b = 0$. In that case, we find only one gapless mode corresponding to the phase mode of the condensate. The second mode is gapped and corresponds to the Higgs or amplitude mode, where the populations at the $A$ and $B$ sublattices oscillate.

In order to investigate the spectrum of collective excitations, we employ the Bogolyubov approach~\cite{Dalfovo-Stringari:1999-rev,Ozeri-Davidson:2005-rev,Pitaevskii-Stringari:book} for the magnon BEC. In this case, one looks for the solution as
\begin{eqnarray}
\label{phen-CM-a-uv}
\psi_{a} &=& \psi_{a,0}(\mathbf{r})+e^{-i \mu t} \left[u_{a}e^{-i\omega t+i\mathbf{k}\cdot\mathbf{r}} +v_{a}^{*} e^{i\omega^* t -i\mathbf{k}\cdot\mathbf{r}}\right],\\
\label{phen-CM-b-uv}
\psi_{b} &=& \psi_{b,0}(\mathbf{r})+e^{-i \mu t} \left[u_{b}e^{-i\omega t+i\mathbf{k}\cdot\mathbf{r}} +v_{b}^{*} e^{i\omega^* t -i\mathbf{k}\cdot\mathbf{r}}\right].
\end{eqnarray}
Here, $\psi_{a,0}(\mathbf{r})$ and $\psi_{b,0}(\mathbf{r})$ denote ground-state solutions, which are, in general, nonuniform. They satisfy the time-independent systems of the Gross-Pitaevskii equations. The perturbations are described by $u_{a,b}$ and $v_{a,b}^{*}$ terms. Finally, $\omega$ is the frequency and $\mathbf{k}$ is the wave vector of collective modes.

\subsection{Ground state}
\label{sec:phenomenology-collective-ground}

Let us start with the ground-state solutions $\psi_{a,0}(\mathbf{r})$ and $\psi_{b,0}(\mathbf{r})$. For simplicity, we consider a uniform ground state. By using Eqs.~(\ref{phen-CM-a-uv}) and (\ref{phen-CM-b-uv}), the Gross-Pitaevskii equations (\ref{phen-GPE-a-def}) and (\ref{phen-GPE-b-def}) lead to
\begin{eqnarray}
\label{PCG-GPE-a}
&&0= \left(c_0-\mu  +\Delta\right) \psi_{a,0} + g_{a} n_{a} \psi_{a,0}  +g_{ab} n_{b} \psi_{a,0} +u_{ab}\psi_{a,0}^{*} (\psi_{b,0})^2,\\
\label{PCG-GPE-b}
&&0 = \left(c_0-\mu -\Delta\right) \psi_{b,0} + g_{b} n_{b} \psi_{b,0}  +g_{ab} n_{a} \psi_{b,0} +u_{ab}^{*}\psi_{b,0}^{*} (\psi_{a,0})^2.
\end{eqnarray}
It is convenient to separate the absolute value and the phase of the ground-state wave functions, i.e., $\psi_{a,0}=\sqrt{n_{a}}e^{i\theta_a}$ and $\psi_{b,0}=\sqrt{n_{b}}e^{i\theta_b}$. As follows from Eqs.~(\ref{PCG-GPE-a}) and (\ref{PCG-GPE-b}), the phase of the ground-state solutions is not fixed for $u_{ab}=0$. Therefore, the state has $U(1)_a\times U(1)_b$ symmetry. If the Josephson coupling term is nonzero, we have
\begin{equation}
\label{phen-coll-J-phase-uab1}
\theta_{u_{ab}}= 2(\theta_{a}-\theta_b) +\pi l, \quad l=0,1,2,3,\ldots,
\end{equation}
where $u_{ab}=|u_{ab}|e^{i\theta_{u_{ab}}}$. Further, we find an effective chemical potential that allows for a solution of the Gross-Pitaevskii equations. We obtain
\begin{eqnarray}
\label{P-C-ground-mu-1}
\mu &=& c_0 +\Delta +g_{a}n_{a} +g_{ab} n_b +u_{ab}n_b,\\
\label{P-C-ground-mu-2}
\mu &=& c_0 -\Delta +g_{b}n_{b} +g_{ab} n_a +u_{ab}n_a.
\end{eqnarray}
Note that even and odd values of $l$ correspond to positive and negative values  of $u_{ab}$ in the equation above.
The system (\ref{P-C-ground-mu-1}) and (\ref{P-C-ground-mu-2}) has a solution for
\begin{equation}
\label{P-C-ground-nb}
n_{b} = \frac{2\Delta +\left(g_{a} -g_{ab} -u_{ab}\right)n_a}{g_{b}-g_{ab} -u_{ab}}.
\end{equation}
(One can fix $n_b$ and solve for $n_a$, which is physically equivalent). The effective chemical potential for the density (\ref{P-C-ground-nb}) reads as
\begin{eqnarray}
\label{P-C-ground-mu-1-nbneq0}
\mu = c_0 +\Delta +g_{a}n_{a} +\left(g_{ab}+u_{ab}\right) \frac{2\Delta +\left(g_{a} -g_{ab} -u_{ab}\right)n_a}{g_{b}-g_{ab} -u_{ab}}.
\end{eqnarray}

There is also an additional solution for Eqs.~(\ref{PCG-GPE-a}) and (\ref{PCG-GPE-b}), where $n_b=0$ and
\begin{equation}
\label{P-C-ground-mu-1-n=0}
\mu = c_0 +\Delta +g_{a}n_{a}.
\end{equation}
We note that solution (\ref{P-C-ground-nb}) exists as long as $n_b>0$. Otherwise, only the ground state with $n_b=0$ and $\mu$ given in Eq.~(\ref{P-C-ground-mu-1-n=0})
is possible.

Therefore, even for simple uniform and static equations for the Dirac BEC, we find two types of solutions: one with nonvanishing magnon densities $n_a\neq0$ and $n_b\neq0$ and the solution that breaks the pseudospin symmetry with $n_a \neq 0$ but $n_b = 0$. The latter solution is an example of nematic instability similar to nematic states in fermion superconductors~\cite{Fradkin-Mackenzie:2010} and in Bose nematics (see, e.g., Ref.~\cite{Jian-Zhai:2011}).

\subsection{Free energy}
\label{sec:phenomenology-collective-F}

In order to clarify which of the two possible solutions for the steady-state homogenous Gross-Pitaevskii equations, discussed in Sec.~\ref{sec:phenomenology-collective-ground}, is the true ground state, we calculate the free energy of the system. By using Eq.~(\ref{phen-F-def}), we derive the following expression:
\begin{eqnarray}
\label{P-C-F-nbneq0}
F_{n_b\neq0} &=& \frac{1}{\left(g_{ab}+u_{ab}-g_{b}\right)^2} \Bigg\{
g_b \left\{n_a^2\left[(g_{ab}+u_{ab})^2+2g_a(g_{ab}+u_{ab})-g_a^2\right] -4g_an_a\Delta -4 \Delta^2\right\} \nonumber\\
&-&n_a(g_{ab}+u_{ab})^2\left[n_a\left(2(g_{ab}+u_{ab})-g_a\right) -4\Delta \right] -g_a g_b^2 n_a^2
\Bigg\},
\end{eqnarray}
where $n_{b}$ given in Eq.~(\ref{P-C-ground-nb}) was used. In the case $n_b=0$, we have a simpler expression
\begin{eqnarray}
\label{P-C-F-n=0}
F_{n_b=0} =-3g_{a} n_a^2.
\end{eqnarray}
In addition to comparing the free energies, one should check whether $n_b$ remains positive for the state with $n_b\neq0$.

\begin{figure*}[!ht]
\centering
\subfigure[]{\includegraphics[height=0.35\textwidth]{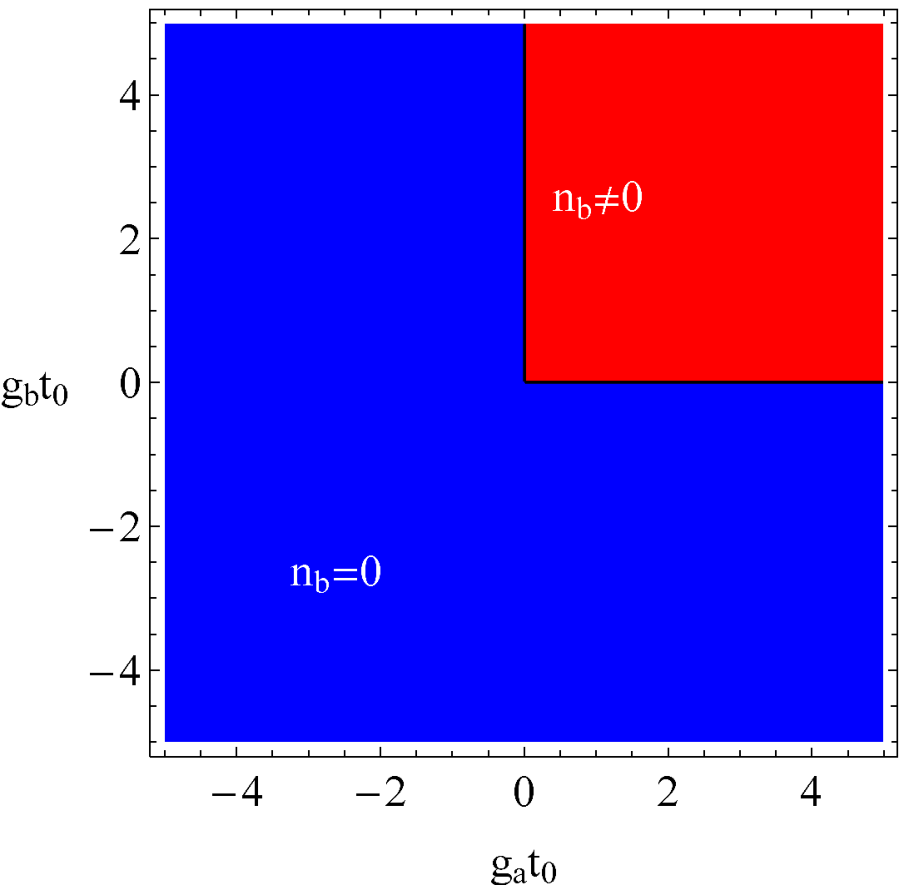}}
\hspace{0.1\textwidth}
\subfigure[]{\includegraphics[height=0.35\textwidth]{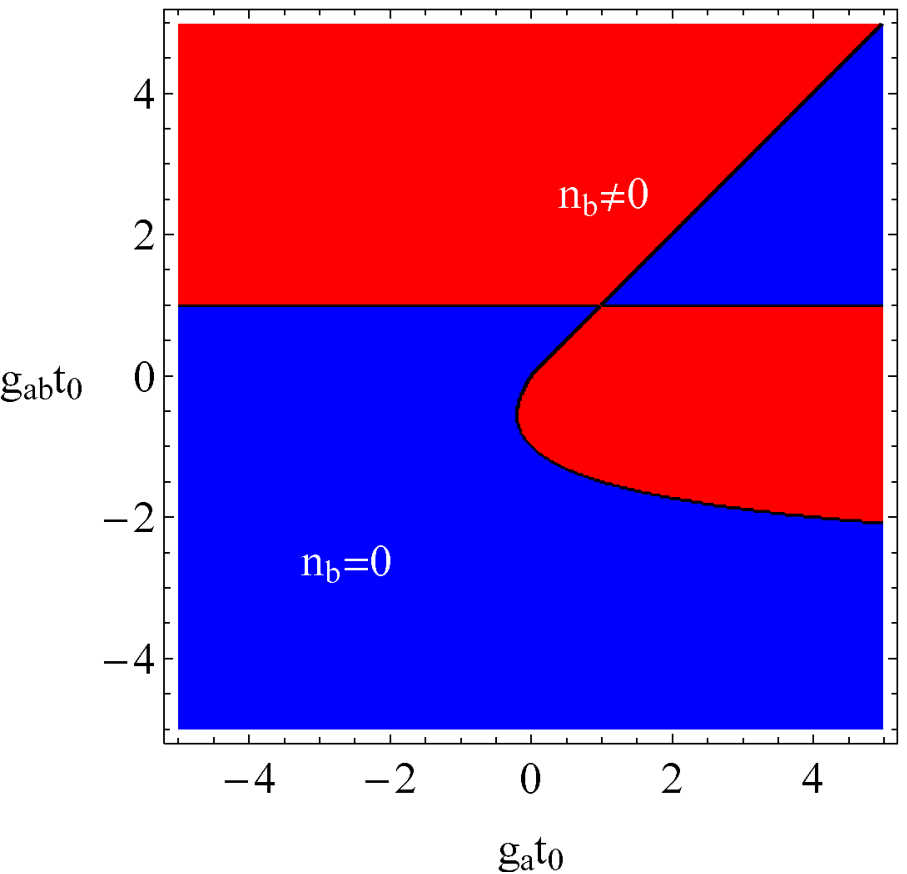}}\\
\subfigure[]{\includegraphics[height=0.35\textwidth]{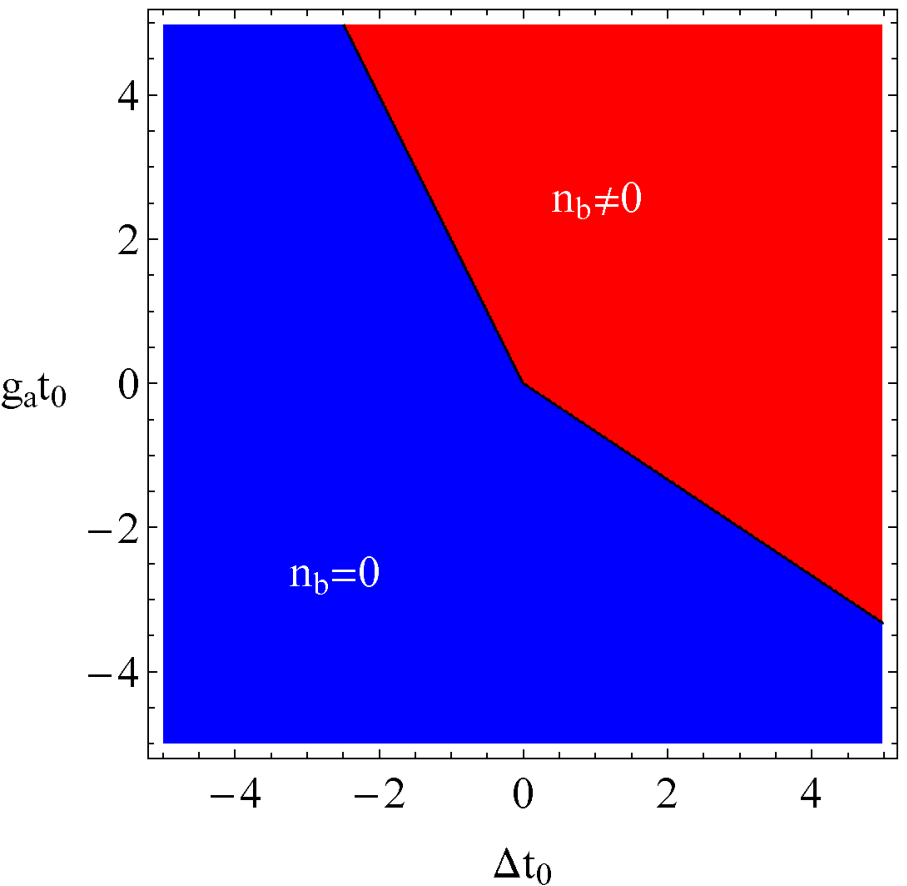}}
\hspace{0.1\textwidth}
\subfigure[]{\includegraphics[height=0.35\textwidth]{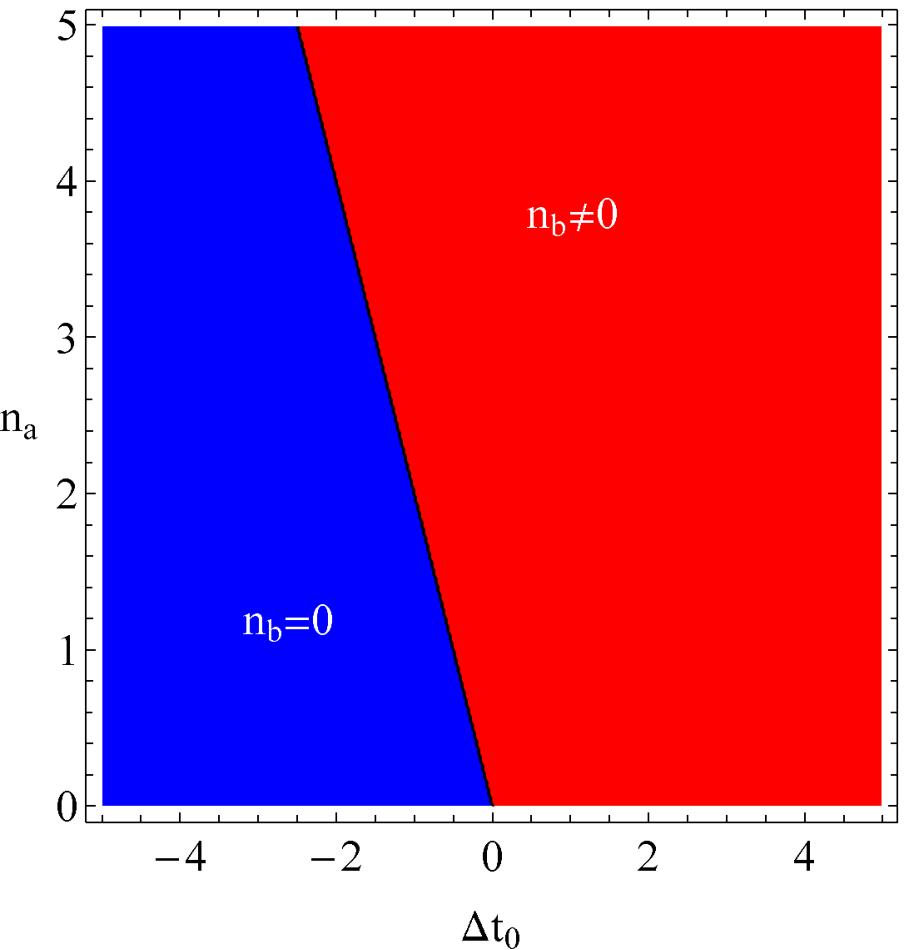}}
\caption{The phase diagram of the system for a few combinations of parameters: $g_{a}$ and $g_{b}$ (panel (a)), $g_a$ and $g_{ab}$ (panel (b)), $\Delta$ and $g_{a}$ (panel (c)), as well as $\Delta$ and $n_{a}$ (panel (d)). Red and blue colors correspond to the ground states with $n_{b}\neq0$ and $n_b=0$, respectively. By default, the values of parameters given in Eqs.~(\ref{phen-three-t-num-vars-be}) and (\ref{phen-three-t-num-vars-ee}) are used.
}
\label{fig:P-C-diff-phase-diagram}
\end{figure*}

We present the phase diagram of the system for a few combinations of parameters in Fig.~\ref{fig:P-C-diff-phase-diagram}. Unless otherwise stated, we use the values of parameters given in Eqs.~(\ref{phen-three-t-num-vars-be}) and (\ref{phen-three-t-num-vars-ee}). The state with the lower free energy is a true (quasi-)ground state. As one can see, the phase diagram is nontrivial where states with both $n_b\neq0$ and $n_b=0$ are possible. In general, however, the ground state with $n_b\neq0$ is favored for repulsive interactions ($g_a>0$ and $g_b>0$). Furthermore, it is clear that the Haldane gap parameter could be used to tune the ground state [see Figs.~\ref{fig:P-C-diff-phase-diagram}(c) and \ref{fig:P-C-diff-phase-diagram}(d)].

\subsection{Dispersion relations}
\label{sec:phenomenology-collective-lin}

In this section, we use the solutions found in Sec.~\ref{sec:phenomenology-collective-ground} and linearize the Gross-Pitaevskii equations (\ref{phen-GPE-a-def}) and (\ref{phen-GPE-b-def}). By using relations (\ref{phen-CM-a-uv}) and (\ref{phen-CM-b-uv}), one obtains the following equations for $u_{a,b}$ and $v_{a,b}$:
\begin{eqnarray}
\label{PCG-GPE-ua}
\omega u_a &=& \left(c_0 +\Delta -\mu\right)u_a +v (k_x-ik_y) u_b +2g_{a} n_a u_a +g_{ab} n_b u_a +g_{ab} \psi_{a,0}\psi_{b,0}^{*}u_b +g_{a} \psi_{a,0}^2 v_a +g_{ab} \psi_{a,0}\psi_{b,0} v_b\nonumber\\
&+& u_{ab} \left[2\psi_{b,0}\psi_{a,0}^{*} u_b +(\psi_{b,0})^2 v_a\right],\\
\label{PCG-GPE-ub}
\omega u_b &=& \left(c_0 -\Delta -\mu\right)u_b +v (k_x+ik_y) u_a +2 g_{b} n_b u_b +g_{ab} n_a u_b +g_{ab} \psi_{b,0}\psi_{a,0}^{*}u_a +g_{b} \psi_{b,0}^2 v_b +g_{ab} \psi_{a,0}\psi_{b,0} v_a\nonumber\\
&+& u_{ab}^{*} \left[2\psi_{a,0}\psi_{b,0}^{*} u_a +(\psi_{a,0})^2 v_b\right],\\
\label{PCG-GPE-va}
-\omega v_a &=& \left(c_0 +\Delta -\mu\right)v_a -v (k_x+ik_y) v_b +2 g_{a} n_a v_a +g_{ab} n_b v_a +g_{ab} \psi_{a,0}^{*}\psi_{b,0}v_b +g_{a} (\psi_{a,0}^{*})^2 u_a +g_{ab} \psi_{a,0}^{*}\psi_{b,0}^{*} u_b\nonumber\\
&+& u_{ab}^{*} \left[2\psi_{b,0}^{*}\psi_{a,0} v_b +(\psi_{b,0}^{*})^2 u_a\right],\\
\label{PCG-GPE-vb}
-\omega v_b &=& \left(c_0 -\Delta -\mu\right)v_b -v (k_x-ik_y) v_a +2 g_{b} n_b v_b +g_{ab} n_a v_b +g_{ab} \psi_{a,0}\psi_{b,0}^{*}v_a +g_{b} (\psi_{b,0}^{*})^2 u_b +g_{ab} \psi_{a,0}^{*}\psi_{b,0}^{*} u_a\nonumber\\
&+& u_{ab} \left[2\psi_{a,0}^{*}\psi_{b,0} v_a +(\psi_{a,0}^{*})^2 u_b\right].
\end{eqnarray}
As before, we omit the valley index $\zeta$ assuming that the valleys have the same dynamics. Note that complex conjugation was performed in Eqs.~(\ref{PCG-GPE-va}) and (\ref{PCG-GPE-vb}).

Equating the determinant of the system (\ref{PCG-GPE-ua}) through (\ref{PCG-GPE-vb}) to zero, the spectrum of collective modes is determined. In what follows, we focus on the case $u_{ab}=0$. The presence of the Josephson coupling term complicates the expressions for the frequencies. For example, numerical calculation suggest that $u_{ab}$ enhances the imaginary parts of the frequencies in the case of the phase-locked condensates ($\theta_a=\theta_b=\theta_{u_{ab}}=0$). For a more detailed discussion and a few dispersion relations at $u_{ab}\neq0$, see Appendix~\ref{sec:App-Josephson-omega}.

In the case of the nontrivial ground state solution given in Eq.~(\ref{P-C-ground-nb}), the frequencies are
\begin{equation}
\label{P-C-lin-omega-u3a=u3b=0}
\omega_{\pm}=\sqrt{v^2k^2 \pm \frac{2v|k_y|}{g_{ab}-g_{b}} \sqrt{\left(g_{ab}-g_{b}\right)\left(g_{ab}^2-g_{a}g_{b}\right)\left[2\Delta +\left(g_{a}-g_{ab}\right)n_a\right]}}.
\end{equation}
Note that both branches are gapless. In addition, we fix the phase of the ground state solutions as $\theta_a=\theta_b=0$. Even at $\Delta=0$, the dispersion relation (\ref{P-C-lin-omega-u3a=u3b=0}) is rather interesting. For example, the mode is stable (i.e., there is no large imaginary part) for $\left(g_{ab}^2-g_{a}g_{b}\right)\left(g_{ab}-g_{a}\right)\left(g_{ab}-g_{b}\right)<0$. In the case $g_a=g_b$, the stability criterion is simple $g_a^2\geq g_{ab}^2$, which agrees with phase-separation criterion for conventional BECs~\cite{Pitaevskii-Stringari:book}.

For the ground state with $n_b=0$, one of the collective modes (e.g., $\omega_+$) can be gapped. The full spectrum in this case is
\begin{eqnarray}
\label{P-C-lin-omega-nb=0}
\omega^2_{\pm} &=& v^2k^2 +\frac{\left[2\Delta +n_a(g_{a}-g_{ab})\right]^2}{2} \nonumber\\
&\pm& \frac{1}{2}\sqrt{\left\{2v^2k^2 +\left[2\Delta +n_a(g_{a}-g_{ab})\right]^2\right\}^2 -4\left\{v^4k^4 +v^2k^2g_{a}n_{a} \left[2\Delta +(g_{a}-g_{ab})n_a\right]\right\}}.
\end{eqnarray}
The gap between $\omega_{+}$ and $\omega_{-}$ branches reads as
\begin{equation}
\label{P-C-lin-omega-gap-nb=0}
\left|\omega_{+}(k=0)-\omega_{-}(k=0)\right| = \left|2\Delta +(g_{b}-g_{ab})n_{a}\right|.
\end{equation}
Note that the gap is determined both by the Haldane gap term and interaction terms. In a simple case $\Delta=0$, the modes at $n_b=0$ are stable at $g_a^2\geq g_{ab}^2$. In addition, while the presence of the gapless modes for 2D materials might naively signify an instability towards long-range power-law correlations, it should not play a crucial role in experimental samples. Indeed, in such a case, the range of correlations is effectively limited by the size of samples and the range of probes.

\begin{figure*}[!ht]
\centering
\subfigure[]{\includegraphics[height=0.35\textwidth]{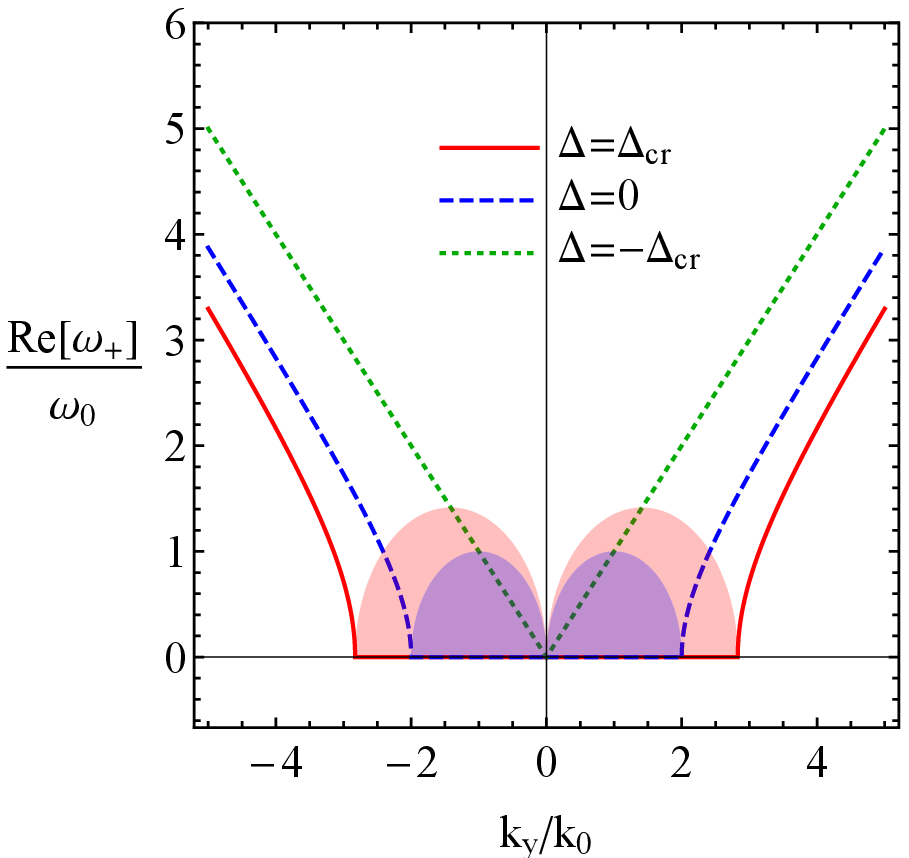}}
\hspace{0.1\textwidth}
\subfigure[]{\includegraphics[height=0.35\textwidth]{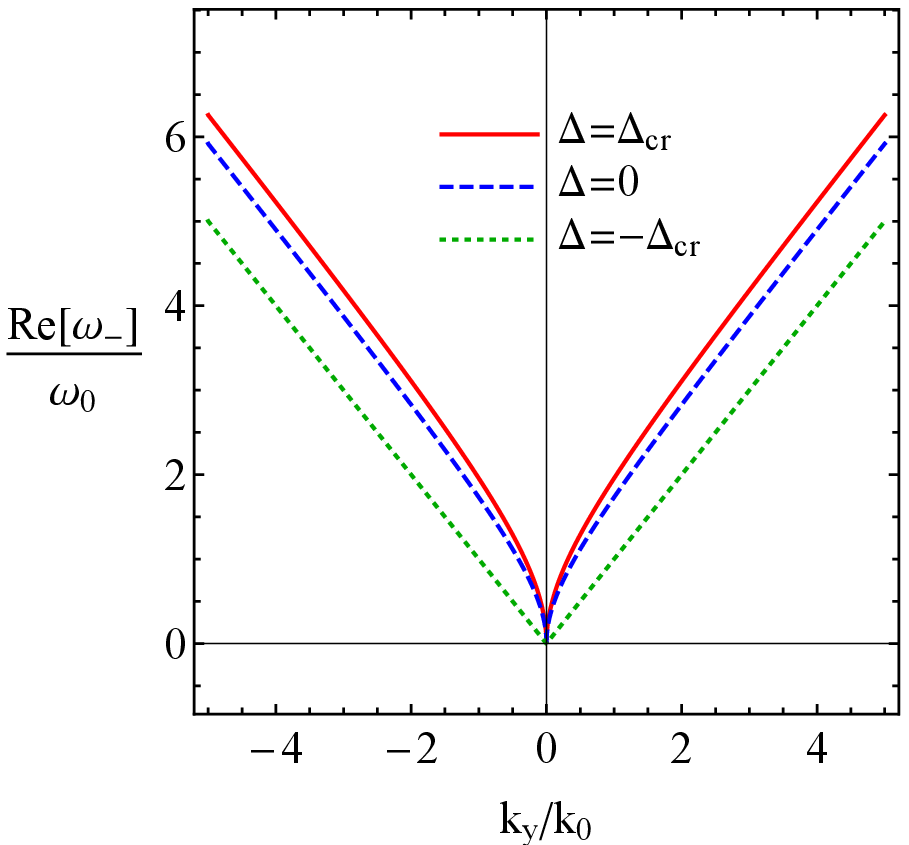}}
\caption{Frequencies of the collective modes $\omega_{+}$ (panel (a)) and $\omega_{-}$ (panel (b)) for $n_b\neq0$ at a few values of the Haldane gap $\Delta$. Here, $k_0=1/(vt_0)$ and $\omega_0=vk_0$. The dispersion relation for $\mathbf{k}\parallel\hat{\mathbf{x}}$ is trivial, i.e., $\omega_{\pm}=vk$. Shaded regions denote the imaginary part of the frequencies. Dispersion relations are gapless for all cases under consideration.}
\label{fig:P-C-lin-sol-1-Delta-2D}
\end{figure*}

\begin{figure*}[!ht]
\centering
\subfigure[]{\includegraphics[height=0.35\textwidth]{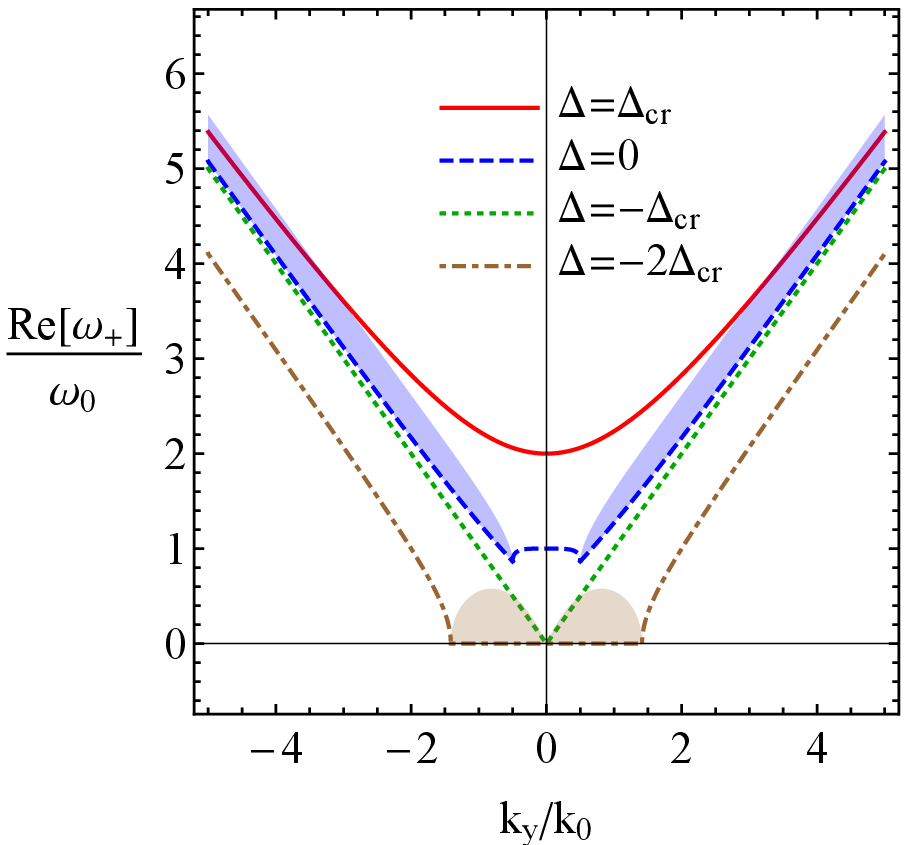}}
\hspace{0.1\textwidth}
\subfigure[]{\includegraphics[height=0.35\textwidth]{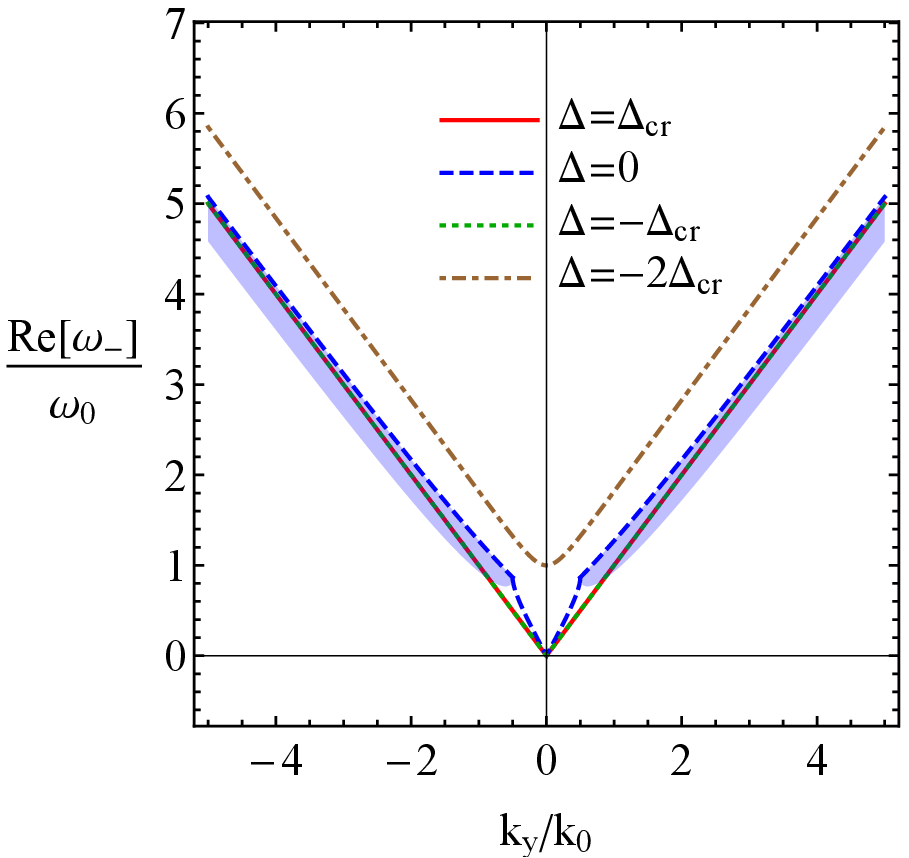}}
\caption{Frequencies of the collective modes $\omega_{+}$ (panel (a)) and $\omega_{-}$ (panel (b)) for $n_b=0$ at a few values of the Haldane gap $\Delta$. Here, $k_0=1/(vt_0)$, $\omega_0=vk_0$, and $\Delta_{\rm cr}=|(g_{b}-g_{ab})n_{a}|/2$. The dependence on $k_y$ is the same as for $k_x$. Shaded regions denote the imaginary part of the frequencies. Except a single point $\Delta =\Delta_{\rm cr}$, one of the modes is gapped and the other is gapless.}
\label{fig:P-C-lin-sol-2-Delta-2D}
\end{figure*}

We present the spectra (\ref{P-C-lin-omega-u3a=u3b=0}) and (\ref{P-C-lin-omega-nb=0}) in Figs.~\ref{fig:P-C-lin-sol-1-Delta-2D} and \ref{fig:P-C-lin-sol-2-Delta-2D}, respectively. It is notable that the rotational symmetry is spontaneously broken in the interacting BEC of magnons at $n_b\neq0$. Indeed, this is evident from the expression in Eq.~(\ref{P-C-lin-omega-u3a=u3b=0}), where an explicit dependence on $k_y$ is present. It can be checked that the anisotropy is set by the relative phase of the ground-state solutions $\psi_{a,0}$ and $\psi_{b,0}$. For example, one could replace $k_y\to k_x$ for $\theta_b-\theta_a=\pi/2$. It is worth noting, however, that the state $n_b\neq0$ is not always the ground state of the system (see the discussion in Sec.~\ref{sec:phenomenology-collective-F}). Depending on the values of the interaction constants, the state with $n_b=0$ could be also realized. Furthermore, both collective modes are gapless. These modes can be identified with two Nambu-Goldstone modes related to the independent oscillations of the condensates at $A$ and $B$ sublattices, i.e., spontaneous breakdown of the symmetries $\psi_{a,0}\to e^{i\theta_a}\psi_{a,0}$ and $\psi_{b,0}\to e^{i\theta_b}\psi_{b,0}$.

The spectrum at $n_b=0$, on the other hand, contains one gapless and one gapped modes. Indeed, the presence of a gapless Nambu-Goldstone mode is guaranteed because $U(1)_a$ symmetry $\psi_{a,0}\to e^{i\theta_a}\psi_{a,0}$ is spontaneously broken in this case. The other mode can be gapped. It corresponds to the Higgs mode of the Dirac BEC. This mode is also known as the amplitude mode because it corresponds to the oscillations of the amplitude of the order parameter. On the other hand, the gapless Nambu-Goldstone mode is related to the oscillations of the phase rather than the absolute value of the BEC's wave function.

Further, we notice that the spectrum of collective modes might contain an imaginary part. The presence of the imaginary part at small $k_y$ for $n_b\neq0$ and large $k$ for $n_b =0$ signifies the dynamical instability of the system. Interestingly, only one of the collective modes (at least at $g_{ab}=0$) has a nonzero imaginary part for $n_b\neq0$. On the other hand, both modes could have a nonvanishing imaginary part at $n_b=0$. Furthermore, we note that the dispersion relation of the collective modes becomes trivial, i.e., $\omega_{\pm}= vk$, at $2\Delta +n_a(g_{a}-g_{ab})=0$. In this case, $n_b=0$.

Let us discuss the dependence of the frequencies of the collective modes on the Haldane gap parameter $\Delta$. As expected from the analytical result (\ref{P-C-lin-omega-u3a=u3b=0}), there is no dependence on $\Delta$ for $\mathbf{k}\parallel\hat{\mathbf{x}}$. On the other hand, the Haldane gap $\Delta$ can be used to tune the dispersion relation of the collective modes at $\mathbf{k}\parallel\hat{\mathbf{y}}$. For example, as one can see from Fig.~\ref{fig:P-C-lin-sol-1-Delta-2D}, one of the modes ($\omega_{+}$) becomes unstable.
The Haldane gap can be also used to stabilize collective modes for $n_b=0$ (see Fig.~\ref{fig:P-C-lin-sol-2-Delta-2D}). In particular, $\omega_{+}$ and $\omega_{-}$ are stable at large values of $\Delta$.

Furthermore, we identify three different phases determined by the combination of parameters $2\Delta +(g_{b}-g_{ab})n_{a}$ at $n_b=0$. For $\Delta<-\Delta_{\rm cr}$, where $\Delta_{\rm cr}=|(g_{b}-g_{ab})n_{a}|/2$, we have one gapped mode and one gapless mode. The latter mode has a nonzero imaginary part and is unstable. These modes merge at $\Delta=\Delta_{\rm cr}$ and have a sound-like dispersion relation $\omega_{\pm}=vk$. One of the modes remains gapless while the other acquires a gap for $-\Delta_{\rm cr}<\Delta<\Delta_{\rm cr}$. The modes are unstable for sufficiently large values of momentum when the spectral branches merge. Finally, the degeneracy is lifted and modes become stable at $\Delta\geq\Delta_{\rm cr}$. For the parameters used, $\omega_{+}$ is gapped and $\omega_{-}$ is gapless in this case. Thus, the Haldane gap appears as an efficient tuning parameter that could be used to access different regimes for collective modes.

\section{Summary and outlook}
\label{sec:Summary}

We investigated the properties of the Bose-Einstein condensate of Dirac quasiparticles and outlined possible routes to achieve this state in bosonic systems. The proposed framework is general and can be applied for investigating various physical systems with Dirac bosonic dispersion. We focus on the case of 2D Dirac magnons. To achieve the Bose-Einstein condensation at the Dirac points, a pumping of magnons was proposed. In particular, we introduced a phenomenological model consisting of the Gross-Pitaevskii equations for the Dirac BEC amended with a rate equation for the pumped magnon population. We found that if the pump power reaches a certain threshold determined by the coupling terms between the pumped and Dirac magnon populations, magnons can populate the Dirac nodes. Depending on the values of the coupling terms, a noticeable depletion might be observed for the pumped magnon populations when the BECs form. The profiles of the magnon densities are nonmonotonic. Indeed, they reach a peak value at the onset. Then, a plateau is developed for a sufficiently large pumping power. Finally, both condensate and pumped magnon population decay when the pump is turned off. The multicomponent condensate coherence is manifested in the Rabi oscillations of the Dirac magnon populations allowed by the Haldane-type gap term and the Josephson coupling. These features of the magnon populations are readily accessible by the Brillouin light scattering technique used for conventional magnons in YIG~\cite{Nowik-Boltyk-Demokritov:2012}. Among other techniques that might be required to resolve magnons belonging to different Dirac nodes or sublattices, we mention time-resolved magneto-optic Kerr effect measurements~\cite{Kruglyak-Grundler:2010} and nitrogen-vacancy centers~\cite{Prananto-An:2020}.

The Dirac nature of magnon BEC is manifested also in the spectrum of collective modes. Even for a simple model at hand, where only the dynamics of a single valley is considered, we found that the spectrum is rather rich. In particular, two uniform (quasi-)ground states are possible, where the magnon density is distributed among the pseudospin (sublattice) degrees of freedom or is accumulated at a certain sublattice. In the first case, the rotation symmetry is spontaneously broken, leading to a directional dependence of the dispersion relation. The direction is determined by the phase of the ground state. We found two branches of gapless collective modes in this case that can be identified with the Nambu-Goldstone modes. Depending on the values of parameters, however, one or even both modes could become unstable for small values of wave vector. The case of a maximally anisotropic distribution of magnons among the sublattices is qualitatively different. In general, the spectrum of one of the branches is gapped and can be identified with the Higgs or amplitude mode. For the certain range of parameters, the modes in this case could be also unstable. It is worth noting that while the case of a maximally anisotropic distribution of magnons seems to be unlikely, it could have lower energy. Independent of the ground state, the dynamics of collective modes could be effectively controlled via the Haldane gap term. This term might be intrinsically present in a system or generated and tuned via the effective pseudo-Zeeman effect in ferrimagnets.

With the results presented in this paper, we hope that the search for the condensates in bosonic Dirac materials will broaden. We reiterate the Dirac equation as a mathematical structure that can be equally applied to fermions and bosons with specific symmetries~\cite{Larsson-Balatsky:2019}. As an explicit example, we considered the case of Dirac magnons, whose experimental realization and condensation would be a new direction to pursue.
Indeed, the majority of the studies in the field of magnon condensates are related to YIG that does not have Dirac points in the magnon spectrum. Trihalides are the promising class of materials that can support Dirac magnons. A suitable class of materials for investigating the Dirac BEC is, however, yet to be identified. We thus expect that the suite of interesting questions about experimental realization, topology, hydrodynamics, edge states, and collective excitations in Dirac BECs would need to be further addressed.

The multicomponent nature of Dirac BEC means that one would have at least four-component hydrodynamics with both normal and superfluid components for the Dirac quasiparticles. Since the condensates are coherent, an interesting problem is the interference of coexisting condensates from different valleys. In a state with different phases of the condensates at different valleys, one would expect spatial interference effects that could be observed by using the Brillouin light scattering, just as for conventional magnon BEC~\cite{Nowik-Boltyk-Demokritov:2012}, as well as time-resolved magneto-optic Kerr effect measurements~\cite{Kruglyak-Grundler:2010} and nitrogen-vacancy centers~\cite{Prananto-An:2020}.
It would be interesting also to investigate the formation of the condensates in finite samples and the effect of the corresponding edge (surface) modes.

Finally, let us discuss a few limitations of this study. The employed model contains a minimal number of terms allowed by the symmetries of the system. Clearly, several other terms could be considered. Among them, we mention the terms corresponding to the inter-valley interactions and nonlinear decay terms. While their effects could be interesting, additional terms greatly expand the parameter space of the problem. Therefore, the corresponding studies will be reported elsewhere. Further, the pumping model proposed in this study is phenomenological. The derivation of the corresponding microscopic model as well as its application to realistic materials is an interesting question that is outside the scope of this study.

\begin{acknowledgments}
We appreciate useful discussions with G.~Aeppli, J.~Bailey, A.~Pertsova, A.~Sinner, C.~Triola, A.~Yakimenko, and V.~Zyuzin. This work was supported by the University of Connecticut, VILLUM FONDEN via the Centre of Excellence for Dirac Materials (Grant No.~11744), the European Research Council under the European Unions Seventh Framework Program Synergy HERO, and the Knut and Alice Wallenberg Foundation KAW 2018.0104. PS acknowledges the support through the Yale Prize Postdoctoral Fellowship in Condensed Matter Theory. SB acknowledges support from Deutsche Forschungsgemeinschaft (DFG, German Research Foundation)-TRR 80.
\end{acknowledgments}

\appendix

\section{Schematic derivation of the energy density for Dirac magnons}
\label{sec:App-magnon-H}

In this appendix, we provide a schematic derivation of the free energy density (\ref{phen-F-def}) by using magnons on a honeycomb lattice as a characteristic example. The Heisenberg Hamiltonian of the model reads as
\begin{eqnarray}
\label{model-H-Heisenberg}
H= -J \sum_{\langle ij\rangle} \mathbf{S}_i \cdot \mathbf{S}_j,
\end{eqnarray}
where the summation runs over the nearest-neighbor coupling between spins $\mathbf{S}_i$. Further, $J$ is the coupling constant, which is positive in the ferromagnetic phase.

The position vectors of lattice sites for the sublattice $A$ could be parameterized as follows:
\begin{equation}
\mathbf{r}_{\mathbf{n}} = n_1 \mathbf{a}_1+n_2 \mathbf{a}_2,
\end{equation}
where the site label $\mathbf{n}=(n_1,n_2)$ is determined by a pair of integers $n_1$ and $n_2$, and
\begin{equation}
\mathbf{a}_1 = \left(\frac{\sqrt{3}a}{2},\frac{3a}{2}\right), \qquad
\mathbf{a}_2 =\left(\frac{\sqrt{3}a}{2},-\frac{3a}{2}\right)
\label{model-primitive-vectors}
\end{equation}
are the primitive translation vectors of the hexagonal lattice and $a$ is the spacing between lattice sites. Further, the position vectors for the sublattice $B$ are given by
\begin{equation}
\mathbf{r}_{\mathbf{n}}^{\prime} = \bm{\delta}_1+ n_1 \mathbf{a}_1+n_2 \mathbf{a}_2,
\end{equation}
where $\bm{\delta}_1 = (\mathbf{a}_1-\mathbf{a}_2)/3$ is the relative position of site $B$ with respect
to site $A$ in the unit cell. The relative positions of the other two sites of type $B$ are given by
$\bm{\delta}_2 = \bm{\delta}_1+\mathbf{a}_2$ and $\bm{\delta}_3 =\bm{\delta}_1-\mathbf{a}_1$.

To bosonise Hamiltonian (\ref{model-H-Heisenberg}), we use the standard Holstein-Primakoff transformation truncated to the first order in the inverse spin $1/S$. By introducing the bosonic annihilation (creation) operators $\hat{a}_i$ and $\hat{b}_i$ ($\hat{a}_i^{\dag}$ and $\hat{b}_i^{\dag}$) corresponding to the two sublattices $A$ and $B$, respectively, we obtain
\begin{eqnarray}
\label{model-HP-def-S1}
S_i^{+} &\equiv& S_i^{x}+iS_i^{y} = \sqrt{2S} \left(\hat{a}_i - \frac{\hat{a}_i^{\dag} \hat{a}_i \hat{a}_i}{4S}\right) +\mathcal{O} \left(S^{-3/2}\right),\\
\label{model-HP-def-S2}
S_i^{-} &\equiv& S_i^{x}-iS_i^{y} = \sqrt{2S} \left(\hat{a}_i^{\dag} - \frac{\hat{a}_i^{\dag} \hat{a}_i^{\dag} \hat{a}_i}{4S}\right) +\mathcal{O} \left(S^{-3/2}\right),\\
\label{model-HP-def-S3}
S_i^{z} &=& S -\hat{a}_i^{\dag}\hat{a}_i.
\end{eqnarray}
Similar expressions can be written for the sublattice $B$. By using these relations and performing the Fourier transform $\hat{a}_i =N^{-1/2} \sum_{\mathbf{k}} e^{i\mathbf{k} \bm{\delta}_i} \hat{a}_{\mathbf{k}}$, where $N$ is the total number of sites and $\mathbf{k}$ is momentum, we derive the following Hamiltonian in the leading order approximation (see also Ref.~\cite{Pershoguba-Balatsky:2018}):
\begin{eqnarray}
\label{model-H-Heisenberg-HP-0}
H= \sum_{\mathbf{k}} \hat{\psi}^{\dag}_{\mathbf{k}} H_0 \hat{\psi}_{\mathbf{k}},
\end{eqnarray}
where
\begin{eqnarray}
\label{model-H0-def}
H_0= JS \left(
          \begin{array}{cc}
            3 & -\gamma_{\mathbf{k}} \\
            -\gamma_{\mathbf{k}}^{*} & 3 \\
          \end{array}
        \right),
\end{eqnarray}
 \begin{eqnarray}
\label{model-Psi-def}
\hat{\psi}_{\mathbf{k}}= \left(
        \begin{array}{c}
          \hat{a}_{\mathbf{k}} \\
          \hat{b}_{\mathbf{k}} \\
        \end{array}
      \right),
\end{eqnarray}
and
\begin{equation}
\label{model-gamma-def}
\gamma_{\mathbf{k}} = \sum_{j} e^{i\mathbf{k}\cdot\bm{\delta}_j} = \left[e^{ia k_y}+2 e^{-\frac{i}{2}a k_y}\cos\left(\frac{\sqrt{3}a k_x}{2}\right) \right].
\end{equation}
The spectrum of Hamiltonian (\ref{model-H0-def}) reads as
\begin{eqnarray}
\label{model-H0-eps}
\epsilon_{\mathbf{k}}^{\pm} = JS\left(3 \pm |\gamma_{\mathbf{k}}|\right) = 3JS \pm JS \sqrt{1+4\cos\left(\frac{\sqrt{3}ak_x}{2}\right)
\left[\cos\left(\frac{\sqrt{3}ak_x}{2}\right)+\cos\left(\frac{3ak_y}{2}\right)\right]}.
\end{eqnarray}

As is easy to check, the two branches in Eq.~(\ref{model-H0-eps}) touch at two non-equivalent isolated points in the Brillouin zone: $\mathbf{k}_{D}^{\pm}= \pm  4\pi/(3\sqrt{3}a) \hat{\mathbf{x}}$, where plus and minus signs correspond to $K$ and $K^{\prime}$ Dirac points or valleys, respectively. Here, $\hat{\mathbf{r}}=\mathbf{r}/|\mathbf{r}|$ is the unit vector. In the vicinity of $K$ and $K^{\prime}$ points, i.e., at $\mathbf{k}=\mathbf{k}_{D}^{\pm}+\delta\mathbf{k}$, where $\delta\mathbf{k}$ is small, the expression for $\epsilon_{\mathbf{k}}^{\pm}$ takes the following simple form:
\begin{equation}
\epsilon_{\mathbf{k}_{D}^{\zeta}+\delta\mathbf{k}}^{\pm}\simeq  
3JS \pm v|\delta \mathbf{k}|,
\end{equation}
where $\zeta=\pm$ is the valley degree of freedom and $v=3JSa/2$ is an analog of the Fermi velocity in graphene. Then, the Hamiltonian in Eq.~(\ref{model-H0-def}) reads as
\begin{eqnarray}
\label{model-H0-lin}
H_0^{(\zeta)}= 3JS \mathds{1}_2 +v\zeta \sigma_x \delta k_x+v \sigma_y \delta k_y.
\end{eqnarray}
Here, $\bm{\sigma} =\left(\sigma_x,\sigma_y\right)$ is the vector of the Pauli matrices acting in the sublattice space. In terms of the bispinor
\begin{equation}
\label{app-model-bispinor-def}
\hat{\Psi}_{\delta \mathbf{k}} = \left(\hat{a}_{+, \delta\mathbf{k}}, \hat{b}_{+, \delta\mathbf{k}}, \hat{b}_{-, \delta\mathbf{k}}, \hat{a}_{-, \delta\mathbf{k}}\right)^{T},
\end{equation}
we have the following Hamiltonian in sublattice and valley space:
\begin{equation}
H_{0} \simeq  \hat{\Psi}_{\delta \mathbf{k}}^\dagger
\left(\begin{array}{cc}
3JS +v (\delta\mathbf{k}\cdot \bm{\sigma})  &  0\\
0 & 3JS -v (\delta\mathbf{k}\cdot \bm{\sigma})
\end{array}\right)
\hat{\Psi}_{\delta\mathbf{k}}.
\label{model-H-4x4-def}
\end{equation}
This magnon Hamiltonian for has a structure of a massless Dirac Hamiltonian. Therefore, we call such quasiparticles \emph{Dirac magnons}.

Further, we add the term describing interaction with an effective magnetic field $\mathbf{B}_i$
\begin{equation}
\label{model-HB-def}
-g\mu_{\rm B} \sum_{i} \left(\mathbf{B}_i\cdot \mathbf{S}_i\right).
\end{equation}
Note that we assume that $\mathbf{B}_i$ could be different at different sublattices. Effectively, this might stem from slightly different gyromagnetic ratios $g$ at the $A$ and $B$ sublattices, which indeed occurs in ferrimagnets~\cite{Fransson-Balatsky:2016}.
It is worth noting also that the pseudo-Zeeman effect was predicted for fermions in graphene in Ref.~\cite{Manes-Vozmediano:2013}. Its origin, however, is related to strains.
For the sake of definiteness, we assume that $\mathbf{B}_i\parallel\hat{\mathbf{z}}$. Performing the Holstein-Primakoff transformation and omitting the constant term $\sim g \mu_{\rm B}BS$, we obtain the following term in the Hamiltonian:
\begin{equation}
\label{model-HB-1}
H_{\rm B} = g\mu_{\rm B} \sum_{\mathbf{{k}}} \hat{\psi}_{\mathbf{{k}}}^{\dag}\left(B \mathds{1}_2 +\tilde{B} \sigma_z\right)\hat{\psi}_{\mathbf{{k}}}.
\end{equation}
As one can see, the usual magnetic field simply shifts the spectrum of magnons as $3JS \to 3JS+g\mu_{\rm B} B$. Therefore, it can be used to tune the effective chemical potential of magnons. Furthermore, it stabilizes the 2D Dirac ferromagnetic material. The effect of the sublattice-dependent field $\tilde{B}=(B_a-B_b)/2$ is qualitatively different. This term opens the gap in the spectrum and is similar to the Zeeman effect albeit in the pseudospin space. Indeed, the energy spectrum (\ref{model-H0-eps}) reads as
\begin{equation}
\label{model-H0-eps-1}
\epsilon_{\mathbf{k}}^{\pm} = JS\left(3 \pm \sqrt{|\gamma_{\mathbf{k}}|^2 +\left(g\mu_{\rm B} \tilde{B}\right)^2/(JS)^2}\right).
\end{equation}
The structure of the effective field $\tilde{B}$ for Dirac magnons allows us to identify it with the Haldane gap term $\Delta$ used in the main text. The energy spectrum of Dirac magnons at $\tilde{B}=0$ and $\tilde{B}\neq0$ is shown in Figs.~\ref{fig:phen-spectrum-schematic}(a) and \ref{fig:phen-spectrum-schematic}(b), respectively.

The next-order Holstein-Primakoff transformation gives the following interaction term~\cite{Pershoguba-Balatsky:2018}:
\begin{eqnarray}
\label{app-model-H4-def}
H_4 &=& \frac{J}{4N} \sum_{\mathbf{k}_1, \mathbf{k}_2, \mathbf{k}_3, \mathbf{k}_4} \delta_{\mathbf{k}_1+\mathbf{k}_2, \mathbf{k}_3+\mathbf{k}_4}\Bigg\{ \gamma_{\mathbf{k}_2}^{*} \hat{a}_{\mathbf{k}_1}^{\dag} \hat{b}_{\mathbf{k}_2}^{\dag} \hat{a}_{\mathbf{k}_3} \hat{a}_{\mathbf{k}_4} + \gamma_{\mathbf{k}_4} \hat{a}_{\mathbf{k}_1}^{\dag} \hat{a}_{\mathbf{k}_2}^{\dag} \hat{a}_{\mathbf{k}_3} \hat{b}_{\mathbf{k}_4}
+\gamma_{\mathbf{k}_2} \hat{b}_{\mathbf{k}_1}^{\dag} \hat{a}_{\mathbf{k}_2}^{\dag} \hat{b}_{\mathbf{k}_3} \hat{b}_{\mathbf{k}_4} +\gamma_{\mathbf{k}_4}^{*} \hat{b}_{\mathbf{k}_1}^{\dag} \hat{b}_{\mathbf{k}_2}^{\dag} \hat{b}_{\mathbf{k}_3} \hat{a}_{\mathbf{k}_4} \nonumber\\
&-&4\gamma_{\mathbf{k}_4-\mathbf{k}_2}^{*} \hat{a}_{\mathbf{k}_1}^{\dag} \hat{b}_{\mathbf{k}_2}^{\dag} \hat{a}_{\mathbf{k}_3} \hat{b}_{\mathbf{k}_4}
\Bigg\}.
\end{eqnarray}
Let us derive the interaction terms in the vicinity of the Dirac points and neglect the gradient terms. This approximation significantly simplifies the final expression. Furthermore, these terms are expected to be weak for small deviations from the Dirac nodes. We use
\begin{eqnarray}
\label{model-gamma-expand-1}
\gamma_{\delta\mathbf{k}} &\approx& 3 +\mathcal{O}(\delta k^2),\\
\label{model-gamma-expand-2}
\gamma_{\mathbf{k}_{D}^{\zeta}+\delta\mathbf{k}} &\approx& -\frac{3}{2}a  \left(\zeta\delta k_x - i \delta k_y\right) +\mathcal{O}(\delta k^2).
\end{eqnarray}
Further, we employ the Fourier transform of the operators
\begin{eqnarray}
\label{EoM-Eq-a-FT-def}
\hat{a}_{\zeta, \delta\mathbf{k}} &=& \frac{1}{\sqrt{N}}\int d\mathbf{r} e^{-i\mathbf{r} \cdot \delta\mathbf{k}} \hat{a}_{\zeta, \mathbf{r}},\\
\label{EoM-Eq-a-FT-1-def}
\hat{a}_{\zeta, \mathbf{r}} &=& \frac{1}{\sqrt{N}}\int \frac{d\delta\mathbf{k}}{(2\pi)^2} e^{i\mathbf{r} \cdot \delta\mathbf{k}} \hat{a}_{\zeta, \delta\mathbf{k}}.
\end{eqnarray}
Note also that
\begin{equation}
\label{EoM-Eq-delta-def}
\delta_{\mathbf{k}_1,\mathbf{k}_2} = \frac{(2\pi)^2}{N} \delta(\mathbf{k}_1-\mathbf{k}_2).
\end{equation}
Then, the leading order interaction term is
\begin{eqnarray}
\label{app-model-H4-fin}
H_4 &\simeq&  -\frac{3J}{N} \sum_{\mathbf{k}_1, \mathbf{k}_2, \mathbf{k}_3, \mathbf{k}_4} \sum_{\zeta_1, \zeta_2, \zeta_3, \zeta_4} \delta_{\zeta_2, \zeta_4} \hat{a}_{\zeta_1, \delta \mathbf{k}_1}^{\dag}\hat{b}_{\zeta_2, \delta \mathbf{k}_2}^{\dag} \hat{a}_{\zeta_3, \delta \mathbf{k}_3} b_{\zeta_4, \delta \mathbf{k}_4} \frac{(2\pi)^2}{N} \int d\mathbf{r} e^{-i\left(\delta \mathbf{k}_1 +\delta \mathbf{k}_2 - \delta \mathbf{k}_4 - \delta \mathbf{k}_3\right) \cdot \mathbf{r}} \nonumber\\
&=& -3J (2\pi)^2\int d\mathbf{r} \sum_{\zeta_1,\zeta_2} \hat{a}_{\zeta_1}^{\dag} \hat{b}_{\zeta_2}^{\dag} \hat{a}_{\zeta_1} \hat{b}_{\zeta_2}.
\end{eqnarray}
Therefore, by using the Bogolyubov approximation for the field operators~\cite{Pitaevskii-Stringari:book}, e.g., $\hat{a}\to\psi_a$, in
\begin{eqnarray}
\label{app-model-H-fin}
H &=& \int d\mathbf{r} \Bigg\{3JS \sum_{\zeta} \left[\hat{a}_{\zeta}^{\dag}\hat{a}_{\zeta} +\hat{b}_{\zeta}^{\dag} \hat{b}_{\zeta}\right] -iv \sum_{\zeta}\hat{a}_{\zeta}^{\dag} \left[\zeta \partial_x -i\partial_y\right]\hat{b}_{\zeta} -iv \sum_{\zeta}\hat{b}_{\zeta}^{\dag} \left[\zeta \partial_x +i\partial_y\right]\hat{a}_{\zeta} -3J(2\pi)^2 \sum_{\zeta_1,\zeta_2} \hat{a}_{\zeta_1}^{\dag} \hat{b}_{\zeta_2}^{\dag} \hat{a}_{\zeta_1} \hat{b}_{\zeta_2}\nonumber\\
&+& \Delta \sum_{\zeta} \left[\hat{a}_{\zeta}^{\dag}\hat{a}_{\zeta} -\hat{b}_{\zeta}^{\dag} \hat{b}_{\zeta}\right]
\Bigg\},
\end{eqnarray}
we obtain the final expression for the energy density
\begin{equation}
\label{EoM-Energy-fin}
\epsilon = c_0 n_{\rm tot} -iv \sum_{\zeta}\psi_{a,\zeta}^{*} \left[\zeta \partial_x -i\partial_y\right]\psi_{b,\zeta} -iv \sum_{\zeta}\psi_{b,\zeta}^{*} \left[\zeta \partial_x +i\partial_y\right]\psi_{a,\zeta} +g_{ab} n_{a}n_{b} +\Delta (n_a -n_b).
\end{equation}
Here, $n_{a,\zeta}=\psi_{a,\zeta}=^{*}\psi_{a,\zeta}=$, $n_{b,\zeta}=\psi_{b,\zeta}^{*}\psi_{b,\zeta}$, $n_a =\sum_{\zeta=\pm}n_{a,\zeta}$, $n_b =\sum_{\zeta=\pm}n_{b,\zeta}$, and $n_{\rm tot} =n_a+n_b$. This energy density serves as a starting point for the free energy in Eq.~(\ref{phen-F-def}). It is worth noting also that we considered a simple model for Dirac magnons in the derivation above. Other interaction terms such as terms corresponding to the Dzyaloshinskii–Moriya and Kitaev interactions, were omitted in this schematic derivation. However, their effects can be still included phenomenologically, at least in part, by adding additional terms to the free energy (\ref{phen-F-def}). The corresponding microscopical analysis will be reported elsewhere.

\section{Classification of non-trivial phases of Dirac BEC}
\label{sec:App-BEC-Class}

In this appendix, we discuss possible non-trivial phases of the Dirac BECs formed at the two valleys of the Brillouin zone. In the absence of interaction, the symmetry class for the BEC corresponds to $U(1)_{a+} \times U(1)_{a-} \times U(1)_{b+} \times U(1)_{b-}$, where each of the symmetries is related to the independent phase rotation of the wave function. We denote the wave-functions $\psi_{\alpha,\zeta}$ ($\alpha = a,b, \zeta = \pm $) at each valley with the pseudospin components as $\psi_{\alpha,\zeta}=\sqrt{n_{\alpha,\zeta}}e^{i\theta_{\alpha,\zeta}}$.
Interaction could lead to phase-locked condensates, a few of which are provided in Table~\ref{tab:Summary}.

\begin{table}[!ht]
\caption{Classification of Dirac condensates residing in two valleys of the spectrum shown in Fig.~\ref{fig:phen-spectrum-schematic} based on their phase locking.
}\label{tab:Summary}
\bigskip
\centering
\setlength{\tabcolsep}{0.7em}
\renewcommand{\arraystretch}{1.6}
\begin{tabular}{c|c|c|c|c}
\cline{1-5} \hline \hline
\multicolumn{1}{c|}{Density} & \multicolumn{1}{c|}{Type} & \multicolumn{2}{c|}{Phase} & \multicolumn{1}{c}{Designation}\\ \hline \hline
\multirow{3}{*}{$n_{a,+} = n_{a,-}$}
& \multirow{3}{*}{Symmetric}
& $\theta_{a,+} = \theta_{a,-}$
& $\theta_{b,+} = \theta_{b,-}$
& Corotating \\ \cline{3-5}
&
& $\theta_{a,+} = -\theta_{a,-}$
& $\theta_{b,+} = -\theta_{b,-}$
& Anti corotating \\ \cline{3-5}
$n_{b,+} = n_{b,-}$
&
& \multirow{2}{*}{$\theta_{a,+} = -\theta_{a,-}$}
& \multirow{2}{*}{$\theta_{b,+} = \theta_{b,-}$}
& Corotating ``B" condensate\\ \cline{1-2}
\multirow{3}{*}{$n_{a,+} \neq n_{a,-}$}
& \multirow{3}{*}{Asymmetric} &
& &  Anti corotating ``A" condensate  \\ \cline{3-5}
&
& \multirow{2}{*}{$\theta_{a,+} = \theta_{a,-}$}
& \multirow{2}{*}{$\theta_{b,+} = -\theta_{b,-}$}
& Corotating ``A" condensate \\
$n_{b,+} \neq n_{b,-}$
& & & & Anti corotating ``B" condensate \\ \hline \hline
\end{tabular}
\end{table}

For simplicity we focus on the possible phase lockings between the condensates at two non-equivalent Dirac points.
The two amplitudes $n_{a}$ and $n_{b}$ at the two different valleys can be equal to each other or might be different. In the former case, we designate the condensates as symmetric and in the latter case, they are asymmetric. For each of these amplitude classes, we then classify the condensates based on the phase relations. If the phases of condensate components are equal in magnitude at the two valleys as $\theta_{a/b,+} = \theta_{a/b,-}$, we call them \textit{corotating}. In the other case, when the phases are equal in magnitude but differ in sign, the condensates are called \textit{anti corotating}. However, for suitable interaction channels between the multicomponent condensates, a completely different phase locking can appear as $\theta_{a,+} = -\theta_{a,-}$ and $\theta_{b,+} = \theta_{b,-}$ (or $\theta_{a,+} = \theta_{a,-}$ and $\theta_{b,+} = -\theta_{b,-}$). In this case, we define it as \textit{anti corotating} ``A" (``B") and \textit{corotating} ``B" (``A").

\section{Effect of intercondensate interactions on the condensate dynamics}
\label{sec:App-BEC-gab}

\begin{figure*}[!ht]
\centering
\subfigure[]{\includegraphics[height=0.27\textwidth]{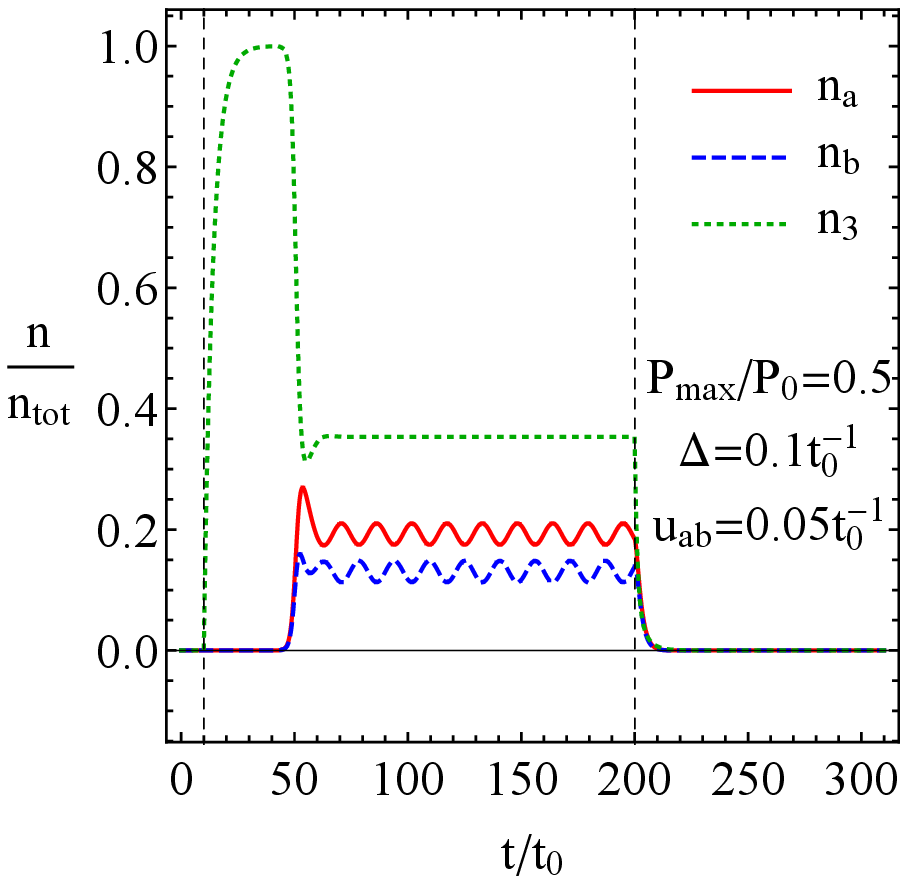}}
\hspace{0.02\textwidth}
\subfigure[]{\includegraphics[height=0.27\textwidth]{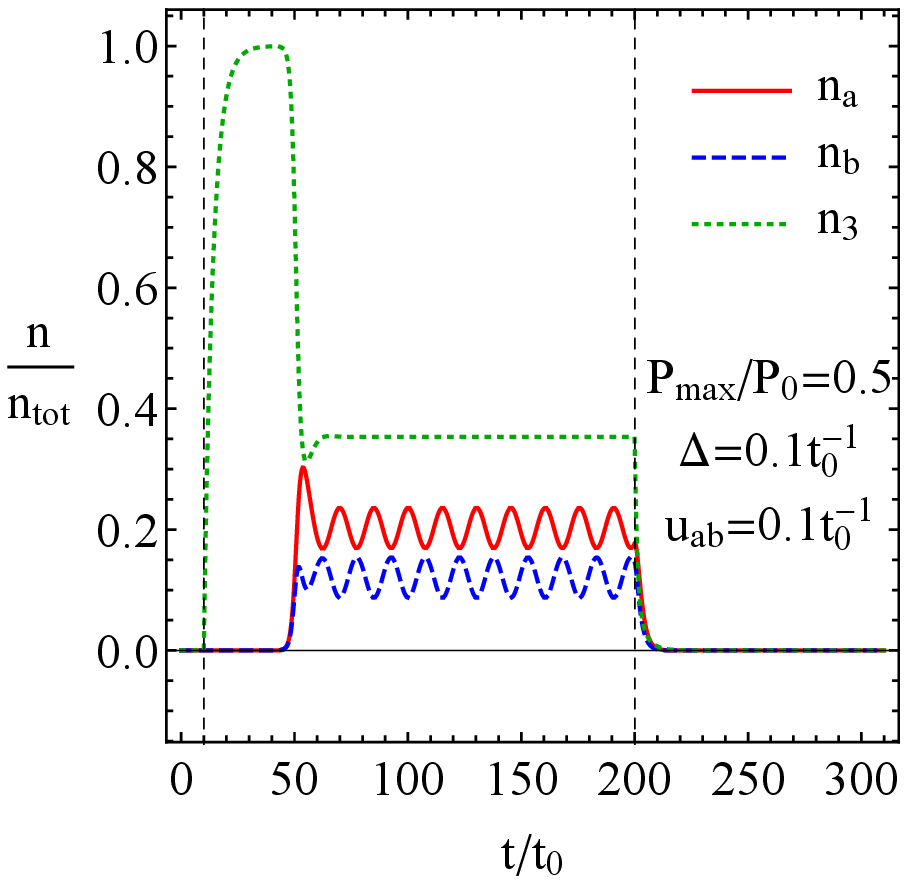}}
\hspace{0.02\textwidth}
\subfigure[]{\includegraphics[height=0.27\textwidth]{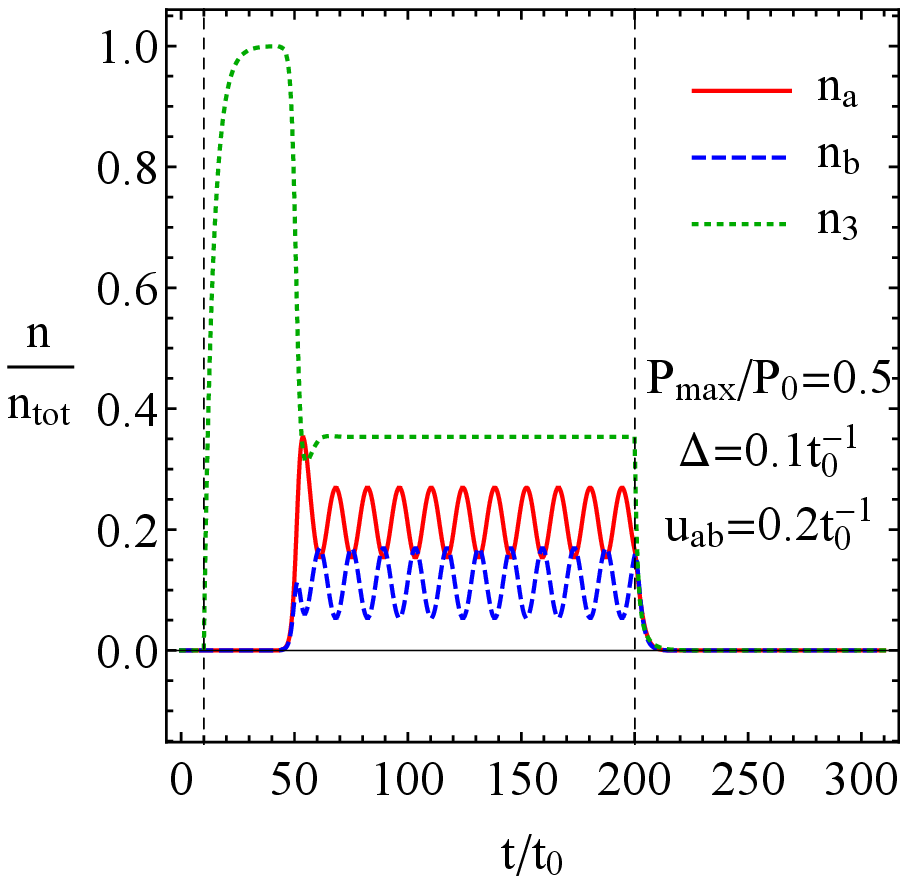}}
\caption{The dependence of the magnon densities on time $t$ at a few values of $u_{ab}$ for $P_{\rm max}=0.5\,P_0$, $\Delta=0.1\,t_0^{-1}$, and $g_a=g_b=g_{ab}=t_0^{-1}$. The densities are normalized to the total magnon density at $t=t_{\rm End}$. Here, the values of other parameters given in Eqs.~(\ref{phen-three-t-num-vars-be}) and (\ref{phen-three-t-num-vars-ee}) are used. In addition, $P_0$ is the characteristic strength of the pump ($P_0=t_0^{-2}$). The Rabi oscillations, whose period is determined by $1/\Delta$ at small $u_{ab}$ and large $\Delta$ ($\Delta>0.1\,t_0^{-1}$ at $u_{ab}=0.2\,t_0^{-1}$), are clearly visible for the magnon BECs.}
\label{fig:App-phen-three-t-P-Josephson-uab1-uab-gab-1}
\end{figure*}

\begin{figure*}[!ht]
\centering
\subfigure[]{\includegraphics[height=0.27\textwidth]{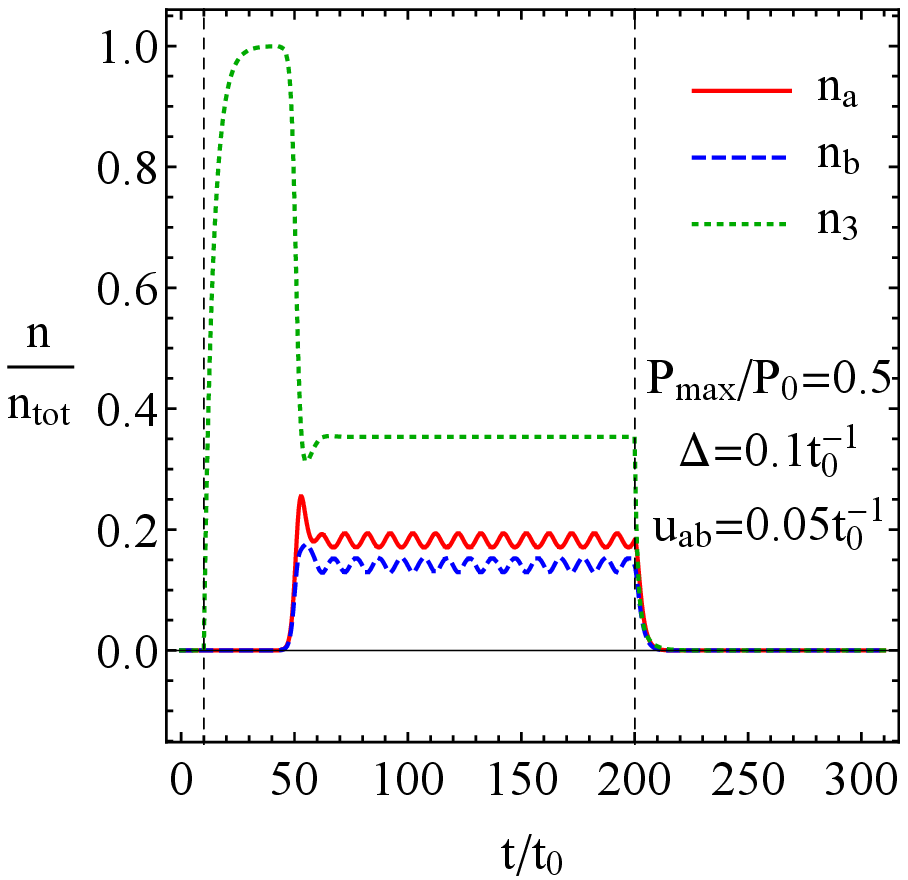}}
\hspace{0.02\textwidth}
\subfigure[]{\includegraphics[height=0.27\textwidth]{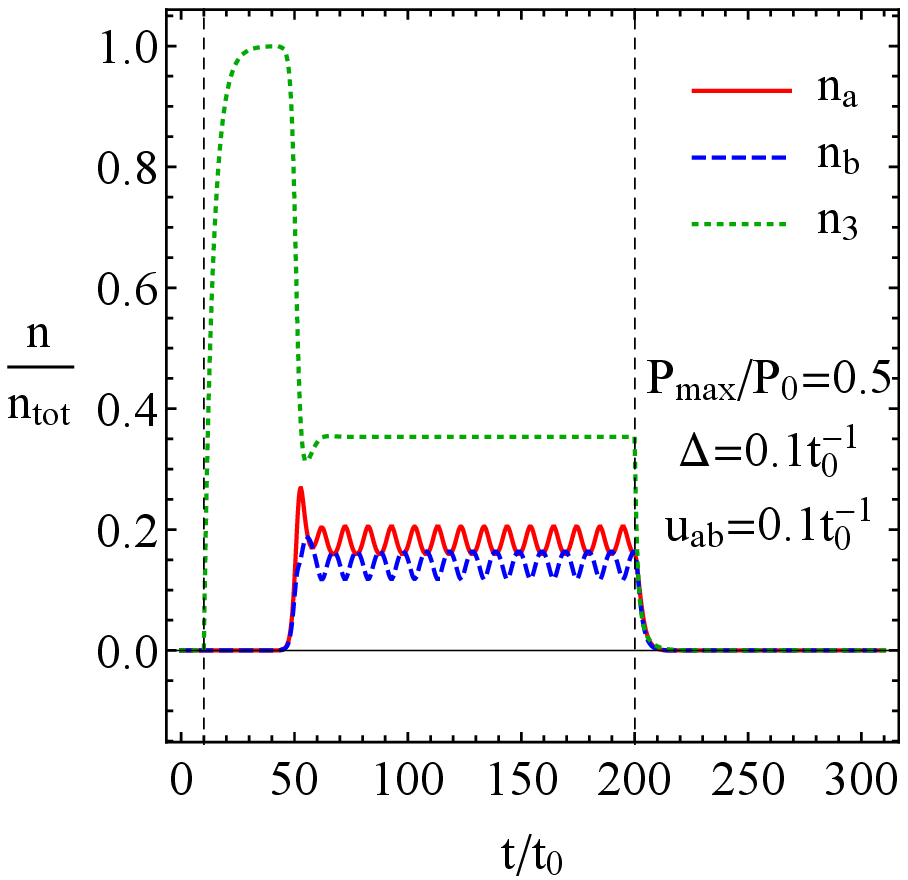}}
\hspace{0.02\textwidth}
\subfigure[]{\includegraphics[height=0.27\textwidth]{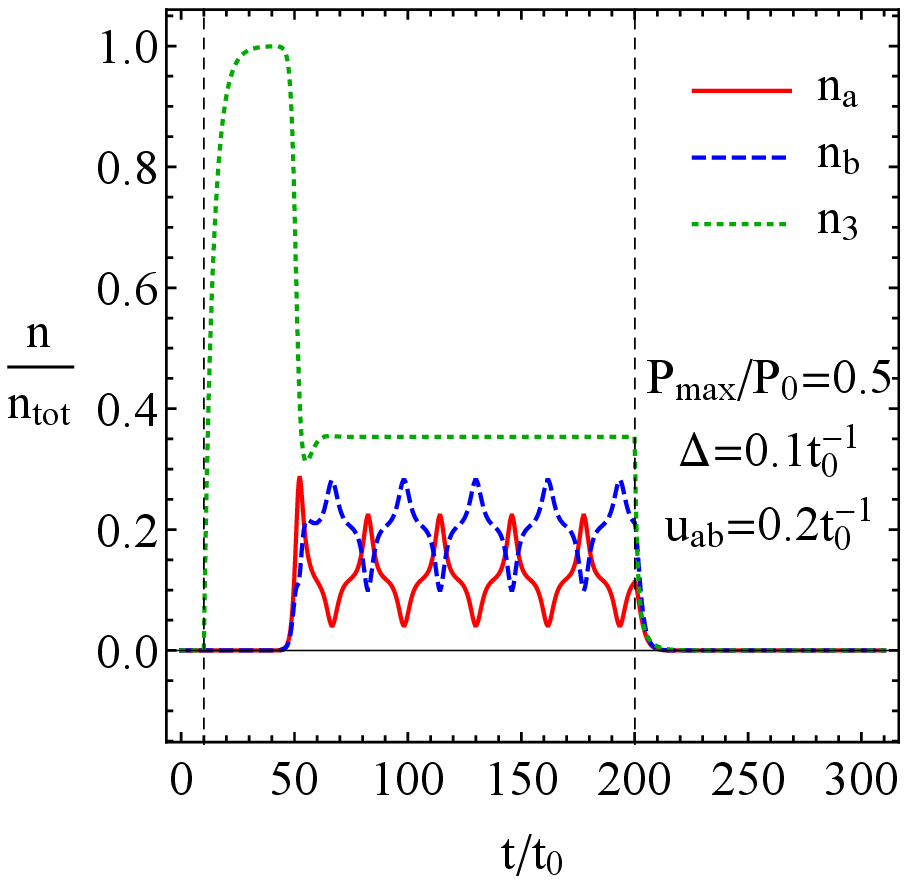}}
\caption{The dependence of the magnon densities on time $t$ at a few values of $u_{ab}$ for $P_{\rm max}=0.5\,P_0$, $\Delta=0.1\,t_0^{-1}$, $g_a=g_b=0$, and $g_{ab}=-t_0^{-1}$. The densities are normalized to the total magnon density at $t=t_{\rm End}$. Here, the values of other parameters given in Eqs.~(\ref{phen-three-t-num-vars-be}) and (\ref{phen-three-t-num-vars-ee}) are used. In addition, $P_0$ is the characteristic strength of the pump ($P_0=t_0^{-2}$). The Rabi oscillations, whose period is determined by $1/\Delta$ at small $u_{ab}$ and large $\Delta$ ($\Delta>0.1\,t_0^{-1}$ at $u_{ab}=0.2\,t_0^{-1}$), are clearly visible for the magnon BECs.}
\label{fig:App-phen-three-t-P-Josephson-uab1-uab-gab-2}
\end{figure*}

Let us consider the effect of nonzero interaction constant $g_{ab}$ on the condensate dynamics. By using the model defined in Sec.~\ref{sec:phenomenology-3component} in the main text, we numerically calculate the magnon densities $n_a$, $n_b$, and $n_3$ for $g_a=g_b=g_{ab}=t_0^{-1}$ (see Fig.~\ref{fig:App-phen-three-t-P-Josephson-uab1-uab-gab-1}) as well as $g_a=g_b=0$, and $g_{ab}=-t_0^{-1}$ (see Fig.~\ref{fig:App-phen-three-t-P-Josephson-uab1-uab-gab-2}). Note that the interaction constants $g_a$, $g_b$, $g_{ab}$ are not manifested in the condensate dynamics when the phase-mixing terms are absent (e.g., the Josephson coupling term $u_{ab}$). As one can see by comparing Fig.~\ref{fig:phen-three-t-P-Josephson-uab1-uab} with  Figs.~\ref{fig:App-phen-three-t-P-Josephson-uab1-uab-gab-1} and \ref{fig:App-phen-three-t-P-Josephson-uab1-uab-gab-2}, the effect of the intercondensate interaction constant $g_{ab}$ is similar to that of $g_{a}$ and $g_{b}$ at small $u_{ab}$. On the other hand, the anharmonic oscillations of the condensates are more easily accessible for a nonzero $g_{ab}$.

\section{Four-component model}
\label{sec:App-phenomenology-4component}

To describe the magnon BECs at the Dirac nodes and at the $\Gamma$ point, we extend the phenomenological model in Sec.~\ref{sec:phenomenology-3component}. In addition, to Eqs.~(\ref{phen-GPE-a-1})--(\ref{phen-GPE-pump-1}), a Gross-Pitaevskii equation for the magnon population at $k=0$ is included. The full system then reads as
\begin{eqnarray}
\label{app-phen-GPE-a}
&&i\partial_t \psi_{a} +i\Gamma_{a}\psi_{a}= \left(c_0-\mu  +\Delta\right) \psi_{a} -iv \left[\left(\partial_x -i\partial_y\right)\psi_{b}\right]
+ g_{a} n_a \psi_{a} +g_{ab} n_{b} \psi_{a} +u_{ab} \psi_{a}^{*}\left(\psi_{b}\right)^{2} \nonumber\\
&&+iP_{a3} \Gamma_{a} n_{3}\psi_{a} -iP_{a0} \Gamma_{0} n_{0}\psi_{a},\\
\label{app-phen-GPE-b}
&&i\partial_t \psi_{b} +i\Gamma_{b}\psi_{b}= \left(c_0-\mu -\Delta\right) \psi_{b} -iv \left[\left(\partial_x +i\partial_y\right)\psi_{a}\right]
+ g_{b} n_b \psi_{b} +g_{ab} n_{a} \psi_{b} + u_{ab}^{*} \psi_{b}^{*}\left(\psi_{a}\right)^{2} \nonumber\\
&&+iP_{b3} \Gamma_{b} n_{3}\psi_{b} -iP_{b0} \Gamma_{0} n_{0}\psi_{b},\\
\label{app-phen-GPE-0}
&&i\partial_t \psi_{0} +i\Gamma_{0}\psi_{0}= -\frac{1}{2m_{\rm eff}} \left(\partial_x^2 +\partial_y^2\right) \psi_{0}
+ g_{0} n_0 \psi_{0} 
+iP_{0b} \Gamma_{a} n_{a}\psi_{0} +iP_{0a} \Gamma_{b} n_{b}\psi_{0},\\
\label{app-phen-GPE-pump}
&&\partial_t n_3 +\Gamma_3 n_3 = \Gamma_3 \tilde{P}(t) -\sum_{\zeta=\pm}\left(P_{3a} \Gamma_{a} n_{a,\zeta} +P_{3b} \Gamma_{b} n_{b,\zeta}\right) n_3. 
\end{eqnarray}
Here, the magnon BEC at the $\Gamma$ point is described by Eq.~(\ref{app-phen-GPE-0}). Due to kinematic constraints, it is unlikely that magnons can scatter to the $\Gamma$ point directly from the pumped population. On the other hand, the lowest branch can acquire magnons via thermalization of magnons at the Dirac points. For example, this could be achieved via four-magnon processes (in the case of YIG, see, e.g., Ref.~\cite{Serga-Hillebrands:2014}).
This process is described via the terms with $P_{0b}$ and $P_{0a}$.

\begin{figure*}[!ht]
\centering
\subfigure[]{\includegraphics[height=0.28\textwidth]{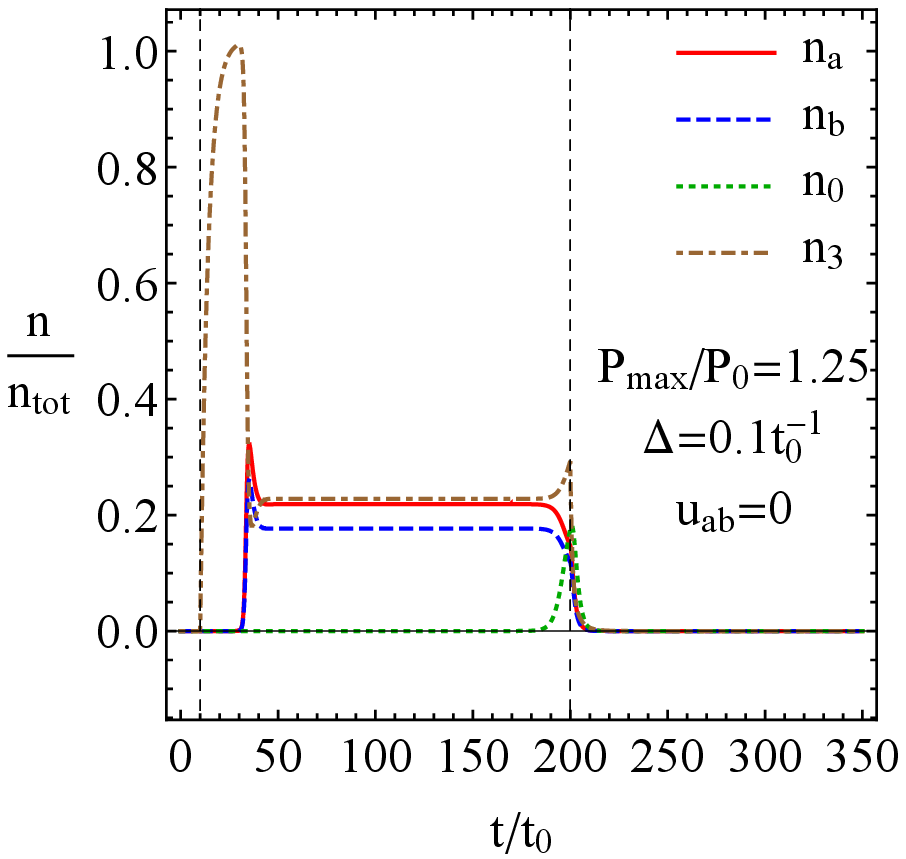}}
\hspace{0.02\textwidth}
\subfigure[]{\includegraphics[height=0.28\textwidth]{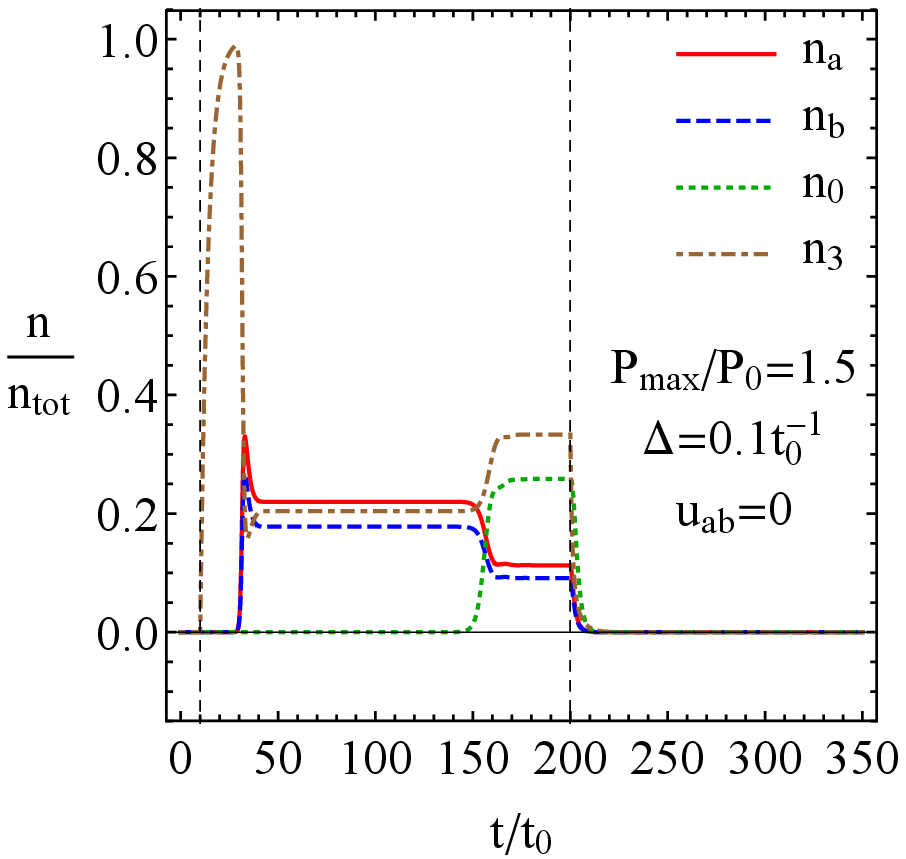}}
\hspace{0.02\textwidth}
\subfigure[]{\includegraphics[height=0.28\textwidth]{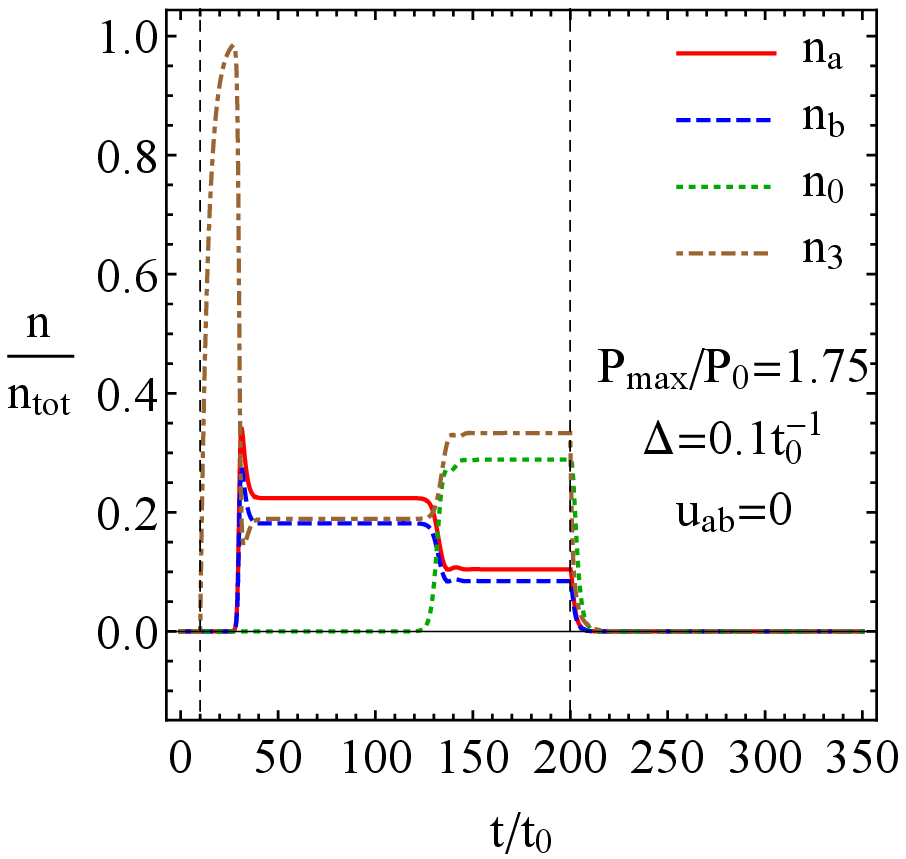}}
\caption{The dependence of the magnon densities on time $t$ at a few values of $P_{\rm max}$ for $\Delta=0.1\,t_0^{-1}$ and $g_a=g_b=g_{ab}=t_0^{-1}$. The densities are normalized to the total magnon density at $t=t_{\rm End}$. Here, $P_0$ is the characteristic strength of the pump ($P_0=t_0^{-2}$).}
\label{fig:phen-four-t-P-Delta-I}
\end{figure*}

By using the values of parameters in Sec.~\ref{sec:phenomenology-time} as well as assuming that $\Gamma_0=\Gamma_a=\Gamma_b=\Gamma_3 = 0.25\,t_0^{-1}$ and $P_{0a}=P_{0b}=0.5$, we present the results for the population densities in Fig.~\ref{fig:phen-four-t-P-Delta-I}. As one can see, the magnon density at the $\Gamma$ point raises at the later stages of the pumping. Since the source of these magnons is the magnon BEC at the Dirac points, the rise of the magnon density at the lowest branch leads to the depletion of the Dirac magnons. Another effect that would be interesting to test experimentally and/or verify microscopically is the rise of the pumped magnon density. In the model at hand, this fact stems from the decrease of the Dirac magnons densities.

\section{Collective modes with Josephson coupling term}
\label{sec:App-Josephson-omega}

\begin{figure*}[!ht]
\centering
\subfigure[]{\includegraphics[height=0.35\textwidth]{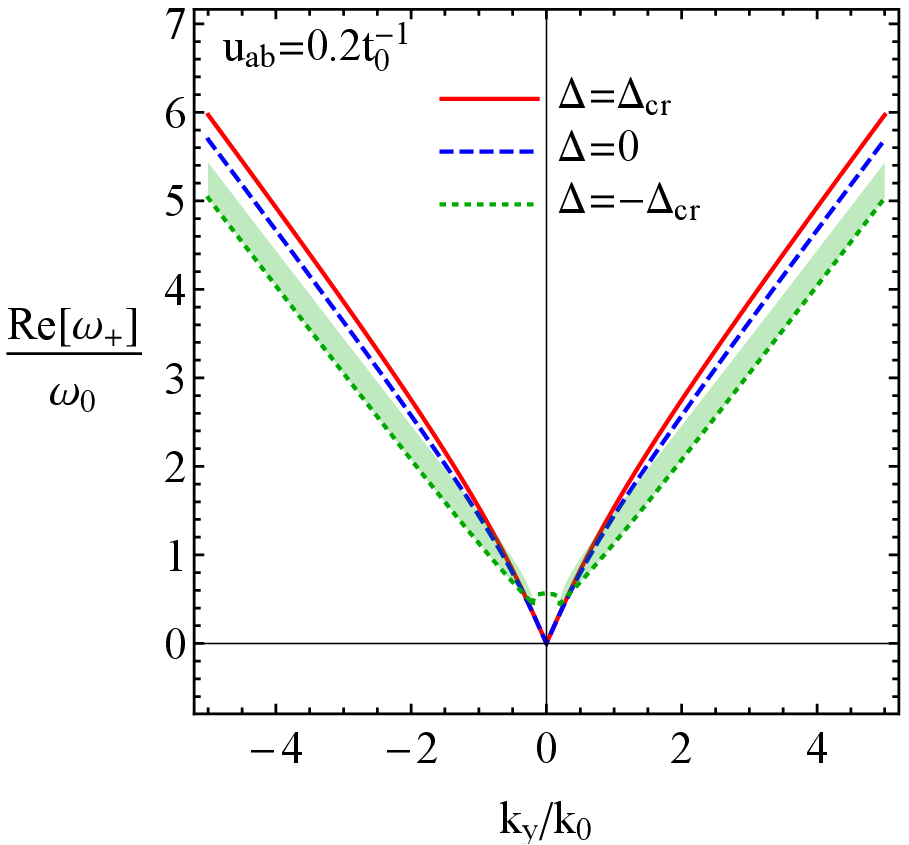}}
\hspace{0.1\textwidth}
\subfigure[]{\includegraphics[height=0.35\textwidth]{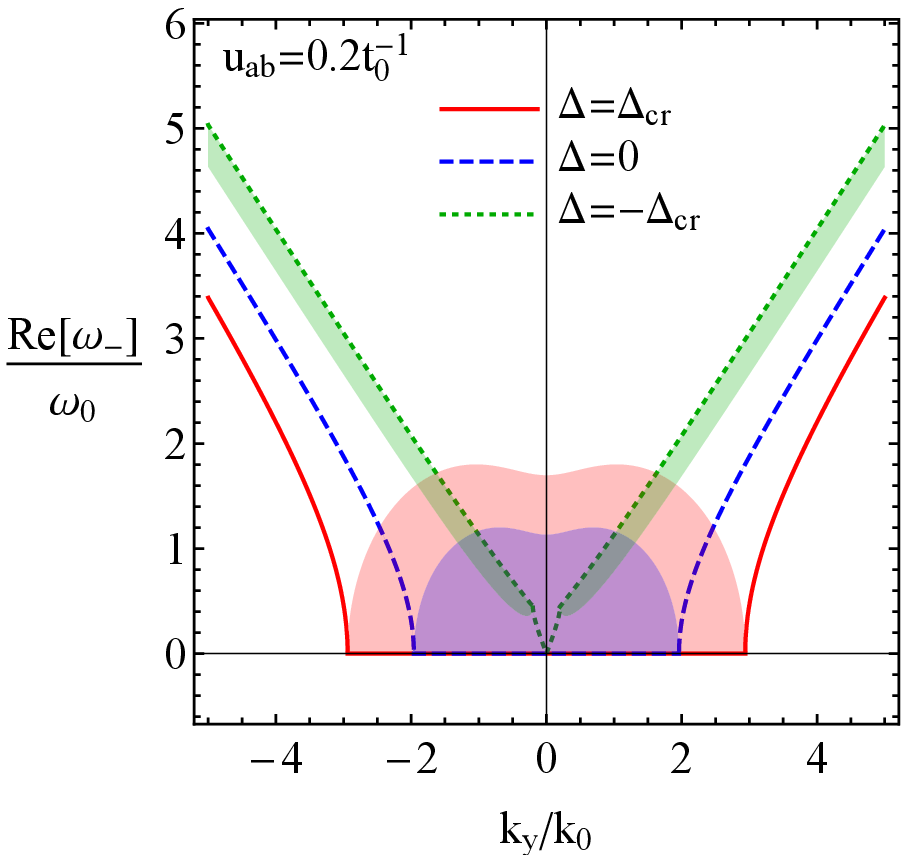}}
\caption{Frequencies of the collective modes $\omega_{+}$ (panel (a)) and $\omega_{-}$ (panel (b)) for $n_b\neq0$ at a few values of the Haldane gap $\Delta$. Here, $k_0=1/(vt_0)$ and $\omega_0=vk_0$. Shaded regions denote the imaginary part of the frequencies.}
\label{fig:app-P-C-lin-sol-1-Delta-2D}
\end{figure*}

\begin{figure*}[!ht]
\centering
\subfigure[]{\includegraphics[height=0.35\textwidth]{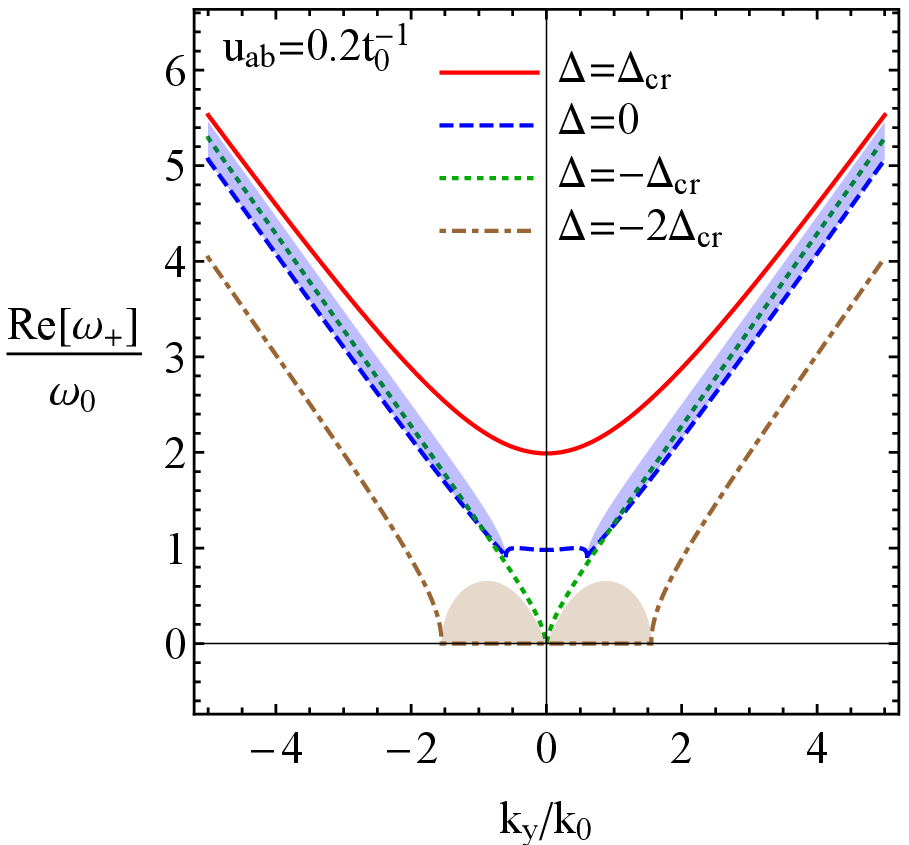}}
\hspace{0.1\textwidth}
\subfigure[]{\includegraphics[height=0.35\textwidth]{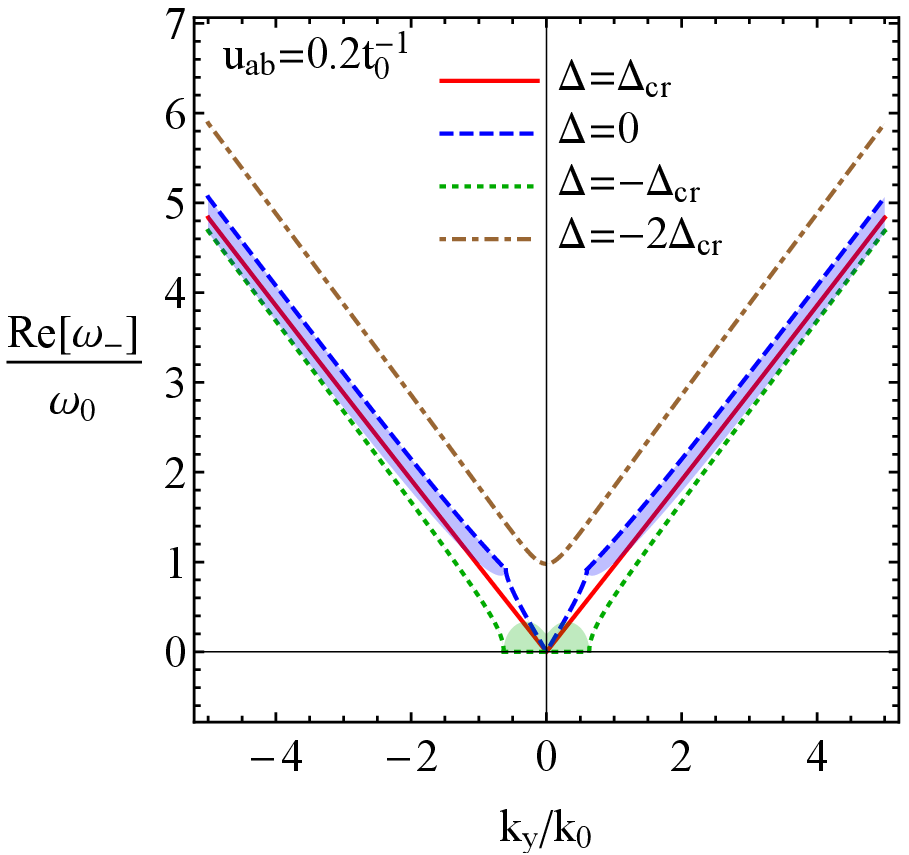}}
\caption{Frequencies of the collective modes $\omega_{+}$ (panel (a)) and $\omega_{-}$ (panel (b)) for $n_b=0$ at a few values of the Haldane gap $\Delta$. Here, $k_0=1/(vt_0)$, $\omega_0=vk_0$, and $\Delta_{\rm cr}=|(g_{b}-g_{ab})n_{a}|/2$. Shaded regions denote the imaginary part of the frequencies. Except a single point $\Delta =\Delta_{\rm cr}$, one of the modes is gapped and the other is gapless.}
\label{fig:app-P-C-lin-sol-2-Delta-2D}
\end{figure*}

Let us briefly discuss the effect of the Josephson coupling term $u_{ab}$ on collective modes considered in Sec.~\ref{sec:phenomenology-collective}. The coupling terms complicate the dynamics of the system and do not allow for a simple analytical solution. Therefore, the frequencies of the collective modes are obtained by solving the system of equations (\ref{PCG-GPE-ua}) to (\ref{PCG-GPE-vb}) numerically with $c_0=g_{a}=g_{b}=t_0^{-1}$, $g_{ab}=0$, and the ground state defined in Sec.~\ref{sec:phenomenology-collective-ground}. The corresponding results for $u_{ab}=0.2\,t_0^{-1}$ are shown in Figs.~\ref{fig:app-P-C-lin-sol-1-Delta-2D} and \ref{fig:app-P-C-lin-sol-2-Delta-2D} at $n_{b}\neq0$ and $n_b=0$, respectively. As one can see, the imaginary part is enhanced compared to the case $u_{ab}$ considered in Sec.~\ref{sec:phenomenology-collective-lin}. Moreover, since we no longer have independent phases for the ground state wave functions at $n_{b}\neq0$, one of the modes can become gapped (see Fig.~\ref{fig:app-P-C-lin-sol-1-Delta-2D}(a)).

\end{document}